%% file: Amanuscript_cr.tex
\documentclass[acmsmall,nonacm]{acmart}
\AtBeginDocument{%
  }

\usepackage{booktabs}
\usepackage{array}
\usepackage{tabularx}
\usepackage{ragged2e}
\usepackage[table]{xcolor}
\usepackage{subcaption}
\usepackage{makecell}

\setcopyright{none}
\renewcommand\footnotetextcopyrightpermission[1]{}
\begin{document}

\title{TibetCPR: A Multimodal Tactile Feedback System to Enhance Cardiopulmonary Resuscitation Training in High-Altitude Regions of Tibet}

\author{Yibo Meng}
\authornote{Yibo Meng and Ruiqi Chen contributed equally to this work.}
\affiliation{%
  \institution{Tsinghua University}
  \city{Beijing}
  \country{China}}
\email{mengyb22@tsinghua.org.cn}

\author{Ruiqi Chen}
\authornotemark[1]
\affiliation{%
  \institution{University of Washington}
  \city{Seattle}
  \state{Washington}
  \country{USA}}
\email{ruiqich@uw.edu}

\author{Zhiming Liu}
\affiliation{%
  \institution{University of Shanghai for Science and Technology}
  \city{Shanghai}
  \country{China}}
\email{10064390@network.rca.ac.uk}

\author{Xiaolan Ding}
\affiliation{%
  \institution{North China University of Science and Technology Health Science Center}
  \city{Tangshan}
  \country{China}}

\renewcommand{\shortauthors}{Meng et al.}

\begin{abstract}
High-quality cardiopulmonary resuscitation (CPR) requires stable control of compression rhythm and depth, yet most training systems presuppose instructor mediation, repeated practice, and explanatory guidance---assumptions that do not hold in the Tibet Autonomous Region, where instruction is fragmented and learners' linguistic and educational backgrounds are heterogeneous. We present \textit{TibetCPR}, a low-cost, self-guided CPR training system that pairs depth-driven electrotactile feedback with rhythm-driven visual cues within a Tibetan-language narrative. In a randomised study with 40 lay community members aged 19--56, the experimental group showed progressive minute-by-minute stabilisation of rhythm and depth across a 10-minute intervention, substantially exceeding an unguided-practice control, with gains transferring to an unscaffolded one-minute post-test. Qualitative accounts described the feedback as legible through participants' bodily action, and usability was high (SUS = 84.3). We synthesise three transferable design principles for self-guided embodied training: feedback as a calibration reference, not an immediate corrector; modality temporal granularity matched to behaviour's temporal structure; and autonomous interpretability as a deployment prerequisite, not an after-effect of usability.

\end{abstract}


\maketitle

\section{INTRODUCTION}
\input{sections/INTRODUCTION}
\section{RELATED WORK}
\input{sections/RELATEDWORK}

\section{SYSTEM DESIGN}
\input{sections/SystemDesign}
\section{METHOD}
\input{sections/ExperimentalDesign}
\section{RESULTS}
\input{sections/Results}
\section{DISCUSSION}
\input{sections/Discussion}
\section{CONCLUSION}
\input{sections/Conclusion}

%\section{GenAI Usage Disclosure}
%In preparing this manuscript, we used ChatGPT in limited ways to support formatting and consistency. Specifically, the tool was employed to assist in generating Overleaf code for the demographic and SUS-content tables, to help transform interview materials into a consistent presentation format, and to check grammar and terminology consistency across sections. The use of AI was restricted to these supportive tasks only. All authors take full responsibility for the content, analysis, and claims made in this paper. All data are real, and all perspectives and interpretations presented are those of the authors; no fabricated data or non-genuine content were produced through the use of large language models.

\bibliographystyle{ACM-Reference-Format}
\bibliography{Aref}

\end{document}

%% file: sections/INTRODUCTION.tex
High-quality cardiopulmonary resuscitation (CPR) is fundamentally an embodied motor skill: rescuers must stabilise compression depth and rhythm continuously through bodily action, often within the first minutes of cardiac arrest \cite{panchal2020part, wong2019epidemiology, grasner2020survival, harris2018cardiopulmonary}. Most existing CPR training systems presuppose that this skill is acquired under instructor mediation, with explanatory guidance, calibrated equipment, and structured rehearsal. Yet in many real-world settings instruction is brief, opportunistic, or absent altogether---and what it means to learn CPR through interaction itself, rather than through interaction with an instructor, remains less well understood.

This question becomes acute in the Tibet Autonomous Region. High altitude and chronic hypoxia raise both the incidence of sudden cardiac events and the physical cost of performing chest compressions, accelerating rescuer fatigue and amplifying the consequences of poor technique \cite{hirota2021hypoxia, sydykov2021pulmonary, yang2022impact, mikolajczak2021impact}. CPR training in the region is at the same time shaped by limited access to qualified instructors, scarcity of simulation equipment, and fragmented opportunities for formal certification \cite{yang2013rapid, yang2024current, wang2022clinicopathological}. CPR learning, when it happens, is short, opportunistic, and entirely self-guided---learners must interpret feedback and adjust their own performance with no external reference frame.

Existing CPR feedback technology has largely not been designed against this constraint. Conventional manikin-based training offers little or no real-time feedback during execution \cite{rao2025enhancing}; high-fidelity simulators provide more precise sensing, but rely on cost, bulk, and expert supervision that are difficult to deploy outside institutional settings \cite{hambly2021rural, lee2021effect, riou2018hijacking}. Both presuppose that feedback will be contextualised by an instructor and that calibration will be guided---a presupposition that does not hold for learners whose access to formal instruction is limited by infrastructure and language.

Serious games have emerged as a partial response, embedding CPR rehearsal in engaging, game-like structures that support repeated practice and self-efficacy across diverse learner populations \cite{khaledi2024comparison, otero2019let}, with deployment flexibility extending across preparatory and independent settings \cite{phungoen2020precourse, drummond2017serious}. However, many existing systems privilege screen-based visual or auditory cues over the embodied force-and-rhythm channel through which CPR is actually executed, leaving open how learners interpret real-time feedback during physically demanding practice---particularly when no instructor is present to disambiguate cues.

In Tibet, these interaction-level challenges are further compounded: extreme environmental conditions can affect the reliability of sensing technologies \cite{rao2025enhancing, raghav2024proportional}, public awareness of bystander CPR remains low \cite{hasselqvist2015early}, and Tibetan-Mandarin linguistic diversity constrains the accessibility of conventional training materials \cite{chen2017public, hooper2011cardiovascular, narahara2012effects}. Rather than treating Tibet as a deployment backdrop, we position it as a context that makes a more general HCI question visible: in self-guided embodied skill training, what design moves let real-time feedback be understood through the user's own bodily action, without external mediation?

To pursue this question, we present \textit{TibetCPR}: a low-cost, self-guided CPR training system organised around three coordinated components---a wrist-worn electrotactile unit, a sensor-instrumented training mannequin, and a game-UI screen running a Tibetan-language narrative. Depth-driven electrotactile signals and rhythm-driven visual cues respond to each compression through deterministic, rule-based mappings, with no instructor mediation. We evaluate the system in a randomised study with 40 lay community members aged 19--56 in the Tibet Autonomous Region, spanning a wide range of occupations and educational backgrounds, none with prior CPR training.

Across both rhythm and depth, performance gains accumulated progressively over the 10-minute intervention rather than at first exposure, and transferred to an unscaffolded one-minute post-test conducted with feedback removed. Drawing together the trajectory data, qualitative accounts, and usability outcomes, we synthesise three design principles for self-guided embodied skill training: (P1) feedback functions as a calibration reference, not an immediate corrector; (P2) modality temporal granularity should match the temporal structure of the target behaviour; and (P3) autonomous interpretability is a deployment prerequisite, not an after-effect of usability. We offer these as transferable design knowledge for mobile, self-guided health training systems beyond institutional settings.

%% file: sections/RELATEDWORK.tex
\subsection{CPR Training Technologies and the Limits of Feedback-Centered Instruction}

CPR training has traditionally relied on physical manikins as the primary instructional medium, reflecting the inherently embodied nature of the task. However, conventional manikin-based systems typically provide limited or no quantitative, real-time feedback on critical performance indicators such as compression depth and rhythm~\cite{rao2025enhancing}. As a result, learners—particularly novices—often depend on indirect cues or post-hoc instructor evaluation to assess performance, making it difficult to identify and correct errors during execution.

To address these limitations, instrumented manikins and advanced CPR training systems have been developed to provide more precise performance measurement. Commercial platforms such as QCPR offer detailed summaries of compression quality and guideline adherence, commonly delivered through post-performance analytics or instructor-facing dashboards \cite{harris2018cardiopulmonary}. High-fidelity simulation systems further extend sensing and evaluation capabilities, but their high cost, physical bulk, and reliance on expert supervision constrain their deployment outside institutional training environments \cite{case2018identifying,meinich2018real,ali2019randomised}. While these systems improve measurement accuracy, their feedback is often delivered after task completion rather than during execution.

Recognizing the need for greater accessibility and independence, recent work has explored more autonomous CPR training approaches that aim to support self-directed learning without continuous instructor oversight \cite{zhang2022current,sevil2021effect,semeraro2011icpr}. Prior studies and systematic reviews of smart-device-mediated CPR training suggest that real-time technology-driven feedback can positively influence skill acquisition outside formally instructor-led sessions \cite{song2016smartwatches, an2019effect}. Nevertheless, many existing systems continue to prioritize the availability and accuracy of feedback signals, while offering limited support for real-time adjustment during physically demanding practice.

These limitations are particularly pronounced in non-institutional and resource-constrained settings, where access to training equipment and professional instruction is limited \cite{liu2021local, an2022training, chen2021needs}. Compounding the infrastructure gap, public awareness of bystander CPR and willingness to act in emergencies remain unevenly distributed across regions and linguistic communities, narrowing the population of users for whom conventional, institutionally framed training is even reachable \cite{chen2017public, teng2020awareness, hasselqvist2015early}. Together, prior work highlights a persistent challenge in CPR training technologies: although feedback has become increasingly precise, it is often delivered in ways that are poorly suited to short, self-guided, and embodied practice scenarios---particularly for users whose access to formal instruction is shaped not only by infrastructure but also by language, literacy, and cultural fit.

\subsection{Serious Games for CPR Training: Engagement and Its Limits}

To address the accessibility and motivational limitations of conventional CPR instruction, serious games have emerged as a widely explored paradigm for emergency care training. By embedding instructional content within game-like structures such as scoring, progression, and immediate feedback, serious games aim to sustain learner engagement while supporting repeated practice \cite{cheng2020part, fijavcko2021evaluating, meng2026engagement, meng2026misty}. Reflecting the growing adoption of this approach, the American Heart Association formally recommends the incorporation of gamified methods into CPR education to enhance instructional quality and learner engagement \cite{cheng2020part}.

Empirical studies and systematic reviews provide strong evidence that game-based CPR training can achieve skill outcomes comparable to traditional instructor-led methods across diverse learner populations, including students and healthcare professionals \cite{cheng2024effects, otero2019let, chen2023evaluation}; broader systematic reviews of serious gaming and gamification in health-professions education report comparable patterns across other clinical skills, suggesting that game-based learning is not narrowly tied to CPR but functions as a more general scaffold for skill rehearsal under self-directed conditions \cite{gentry2019serious, chen2023design}. In addition to technical performance, prior work has reported positive effects on learners’ confidence and self-efficacy, which are widely recognized as important factors for willingness to intervene in real-world cardiac arrest situations \cite{khaledi2024comparison}.

Despite these advantages, many serious game–based CPR systems simplify aspects of embodied skill execution. A substantial portion of existing designs rely primarily on visual or auditory feedback delivered through screens, with limited integration of physical sensing or haptic guidance \cite{phungoen2020precourse, de2019comparative}. While such approaches lower barriers to deployment and support scalability, they often foreground symbolic performance representations—such as scores, indicators, or virtual outcomes—over continuous bodily engagement during chest compressions.

Prior research has noted that this emphasis may introduce challenges for novice learners and self-guided training contexts, particularly when precise force control and rhythm regulation are required \cite{drummond2017serious,de2019comparative}. These observations suggest a tension between the motivational strengths of serious games and their ability to support physically grounded skill acquisition in autonomous practice scenarios.

Together, existing work establishes serious games as a compelling and effective foundation for CPR education, while also highlighting open challenges in aligning engagement-driven designs with the demands of embodied motor learning. These limitations motivate further investigation into how game-based CPR systems can better support physical skill development beyond screen-based interaction alone \cite{chen2025gestobrush, zhang2025openhoiopenworldhandobjectinteraction}.

\subsection{Multimodal and Haptic Feedback for Embodied Skill Learning}

High-quality CPR is fundamentally an embodied motor skill that requires coordination of force, rhythm, and bodily posture. Prior work in HCI and health informatics has shown that reliance on indirect visual or auditory cues alone is often insufficient for developing stable compression techniques, particularly for novice learners who lack prior physical reference points against which to calibrate their own action \cite{abella2005quality, zhao2025immersive, cao2022multichannel}. This observation aligns with a broader account in CPR feedback research in which real-time signals support iterative self-adjustment of force and timing rather than dictating moment-to-moment correction \cite{kramer2006quality, song2016smartwatches}---a framing that becomes especially consequential when training takes place without instructor mediation.

To better support embodied learning, CPR training technologies have increasingly explored the use of haptic feedback. Existing approaches broadly include instrumented manikins and wearable feedback devices. Force-sensitive manikins can provide detailed performance analytics, but their feedback is often delivered after task completion, limiting support for real-time motor adjustment during practice \cite{tanaka2019effect,rauzi2024current}. Wearable systems, such as sensor-equipped gloves and other body-worn devices, offer closer temporal coupling between action and feedback, but frequently rely on a single feedback modality \cite{musiari2021can,guridi2024proof}.

Within the broader HCI literature, different haptic modalities have been shown to afford distinct perceptual characteristics. Vibrotactile feedback is generally reported as intuitive and well accepted, while thermal and electrotactile feedback can increase perceptual salience \cite{lee2024soft,moorthy2024attributes,lu2020wearable}. However, most existing CPR training systems employ these modalities in isolation, and the comparative role of different haptic signals—and their potential combination within a single training system—remains underexplored \cite{lin2025use,swain2024assessing}.

Beyond CPR specifically, recent work on electrotactile and audio-tactile feedback in adjacent motor-learning domains---including finger posture training, upper-extremity coordination, and proprioceptive recalibration---suggests that brief, error-locked tactile signals can support iterative motor adjustment with relatively low continuous attentional cost \cite{vargas2008audio, doan2025electro, ravichandran2024electrotactile}. Together with the modality-comparison work above, these results hint that the temporal structure of tactile feedback---continuous tracking versus discrete error signalling---may matter as much for self-guided motor learning as the choice of modality itself. Within CPR training research, however, this temporal dimension has rarely been examined, and the question of how it interacts with the structure of the underlying control variables (continuous rhythm versus categorical depth) remains open.

A further limitation concerns deployment context. Many haptic CPR training systems are evaluated under controlled laboratory conditions and assume access to expert supervision or structured training sessions \cite{kahsay2026comparison,lins2022evolutionary}. As a result, less is known about how multimodal haptic feedback supports learning in short, self-guided practice scenarios, where users must interpret and respond to feedback without instructor mediation.

Taken together, prior work highlights the promise of haptic feedback for supporting embodied CPR learning, while leaving open questions about modality integration and suitability for autonomous, real-world use. These gaps motivate further investigation into how multimodal feedback can be incorporated into CPR training systems designed for self-guided practice beyond laboratory settings.

%% file: sections/SystemDesign.tex
\subsection{Design Goals and Contextual Constraints}

The deployment context outlined in Section~1---chronic hypoxia, sparse emergency-medical infrastructure, and the near-absence of structured CPR curricula across the Tibetan Plateau---invalidates several assumptions that are routinely embedded in conventional CPR training systems. First, repeated instructor-guided calibration cannot be assumed; learners must rely on the system itself to establish and correct motor boundaries. Second, long-duration or intensive training sessions are impractical because physical fatigue is amplified by reduced oxygen availability. Third, continuous or densely multimodal feedback risks overwhelming users during physically demanding compressions, particularly in the absence of expert mediation to disambiguate signals. Finally, motivation to return to training cannot rely on institutional requirements or certification pathways and must instead be sustained intrinsically.

TibetCPR responds to these constraints with four design commitments. The system prioritises short, self-guided training encounters; salient but intermittent feedback rather than continuous stimulation; boundary-based correction in place of dense procedural instruction; and a longer-arc motivational scaffold that does not depend on external accountability.

At the level of feedback, these commitments are realised through two channels paired to match the temporal structure of the variable each is asked to support: a continuous visual rhythm channel for the continuously-regulated compression rate, and a discrete electrotactile depth channel that fires only when a compression falls outside the recommended depth range. This division of labour is grounded in motor-learning evidence. Augmented multimodal feedback supports motor-skill acquisition more effectively than single-modality cues \cite{sigrist2013augmented}; electrotactile signalling has been shown to produce faster learning rates than no-feedback baselines \cite{villa2025understanding}; sensor-driven visual feedback supports motor coordination in self-directed contexts \cite{hegi2023sensor}; and corrective-feedback devices show measurable benefits for CPR skill acquisition specifically \cite{nicolau2024influence, lin2025use}.

At the level of cultural and contextual fit, the design responds to deployment realities at the population level. Public CPR awareness in the region is low, learners' linguistic and educational backgrounds are highly heterogeneous, and many users are not primarily Mandarin-readers. The snow-lotus narrative, Tibetan-language prompts, and highland-pasture visual register are chosen so that the system's evaluative signals can be acted on through bodily interaction with minimal reliance on text. The low hardware cost reflects the same logic: in a region where training equipment is rarely sustained at scale, a system that can be reproduced at low cost is more likely to actually reach users.

Together, these commitments and design choices shape the system architecture (Section~3.2), the narrative game layer (Section~3.3), and the multimodal feedback design (Section~3.4) presented below.

\subsection{System Overview}

Guided by these constraints, TibetCPR is composed of three coordinated physical components: a wrist-worn unit, a standard CPR training mannequin, and a screen running the game UI. The wrist-worn unit is a fingerless glove housing a microcontroller and a compact OLED status display, with two lead wires terminating in hydrogel electrodes that adhere to the user's wrist and deliver electro-tactile stimulation. The mannequin embeds a piezoelectric pressure sensor in its torso that records compression peak force and timing throughout each compression. The game-UI screen, positioned within the user's line of sight during practice, runs the narrative game layer described in Section~3.3, and the hardware components are detailed and depicted together in Section~3.4. Physical chest compression on the mannequin is the sole and primary control input, and all system responses are driven directly by the user's real-time compression performance.

Within each compression cycle, the system runs a single deterministic loop. The mannequin's pressure sensor reports the peak force and timing of every compression to the wrist-worn microcontroller over a short wired serial link, and the microcontroller in turn pushes per-compression state updates to the game-UI host over a Bluetooth Low Energy link. On each update the microcontroller evaluates two independent rules in parallel: a depth rule that maps the compression's peak force to one of three electro-tactile outputs (TENS, none, or EMS; see Section~3.4), and a rhythm rule that maps the running compression rate to one of three colour states (blue, green, or red) on the game-UI display. Game-state progression---the gradual blooming of the snow lotus described in Section~3.3---is updated in synchrony with successful compressions, allowing the system to function simultaneously as moment-to-moment correction support and as a longer-arc motivational scaffold in the absence of an instructor.

\subsection{Game and Narrative Layer}

\begin{figure}[t]
  \centering
  \includegraphics[width=0.57\linewidth]{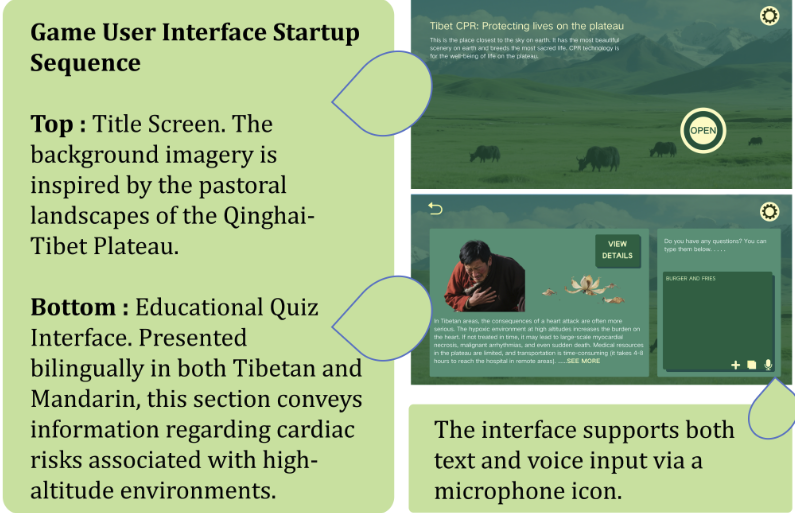}
  \caption{Game-UI start sequence: title screen (\textit{top}) and the bilingual Tibetan/Mandarin educational screen (\textit{bottom}). Voice input via the microphone icon is provided to accommodate participants with limited literacy.}
  \label{fig:interface}
\end{figure}

The game layer running on the game-UI screen serves three intertwined purposes: it provides cultural framing that increases willingness to engage among local users, supplies a longer-arc narrative progression that sustains attention through the inherently repetitive motor task, and lowers the educational-access threshold for users with limited prior exposure to emergency medicine. The interface is set in a Tibetan highland pasture---yaks grazing under snow-capped mountains---a culturally familiar scene chosen so that the visual environment of the system reads as part of participants' own surroundings (Fig.~\ref{fig:interface}).

Sessions begin with a title screen (Fig.~\ref{fig:interface}, top) and an educational/Q\&A screen (Fig.~\ref{fig:interface}, bottom) in which short text---rendered in Tibetan and Mandarin---explains the heightened risks of cardiac emergencies at high altitude, the long pre-hospital intervals typical of remote Tibetan regions, and the rationale for early bystander response. A withered snow lotus, a revered Tibetan symbol of resilience, accompanies this text and establishes an emotional objective: to revive a fading life. The same screen exposes a chat interface that accepts both typed and voice input (microphone icon), allowing users to ask follow-up questions about the educational content---a feature included specifically to accommodate participants with limited literacy.

\begin{figure}[t]
  \centering
  \includegraphics[width=0.7\linewidth]{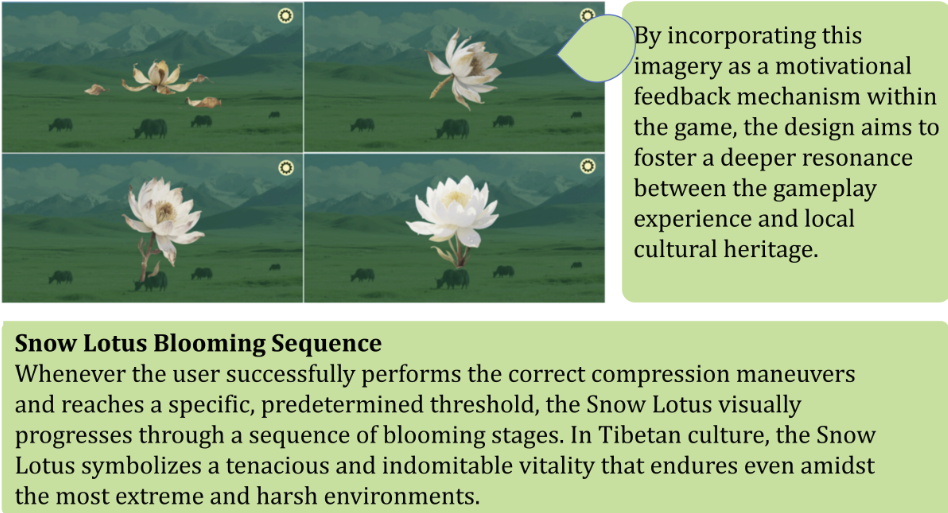}
  \caption{Snow-lotus blooming sequence used as the in-game progress visualisation.}
  \label{fig:snow-lotus}
\end{figure}

Once practice begins, gameplay unfolds as a continuous interactive experience in which physical CPR compressions on the mannequin directly drive on-screen progression. The snow lotus serves as the central progress indicator: with each correctly executed compression, the lotus advances through a four-stage blooming sequence---from withered, to partial bloom, to near-full bloom, and finally to full bloom (Fig.~\ref{fig:snow-lotus})---that constitutes the game's success visualisation. Incorrect compressions---those falling outside the recommended depth range---do not trigger any narrative penalty within the lotus visual; instead, they trigger the electro-tactile depth signal described in Section~3.4, which is narratively framed as a tangible warning to refocus attention.

The snow lotus was chosen as the progress visualisation for two reasons. First, it is a recognised Tibetan cultural symbol associated with tenacious life under harsh natural conditions, providing a visual register that is locally meaningful in the region we target. Second, a slowly-blooming flower whose final state is reached only after sustained practice supplies a longer-arc accumulation cue than a tick-by-tick score, supporting motivation across the repetitive compression task when training takes place without instructor follow-up. Both choices are echoed in participants' interview accounts (Sections 5.3.2 and 5.3.4).

\subsection{Hardware Architecture and Multimodal Feedback}

\begin{figure}[htbp]
  \centering
  \includegraphics[width=\linewidth]{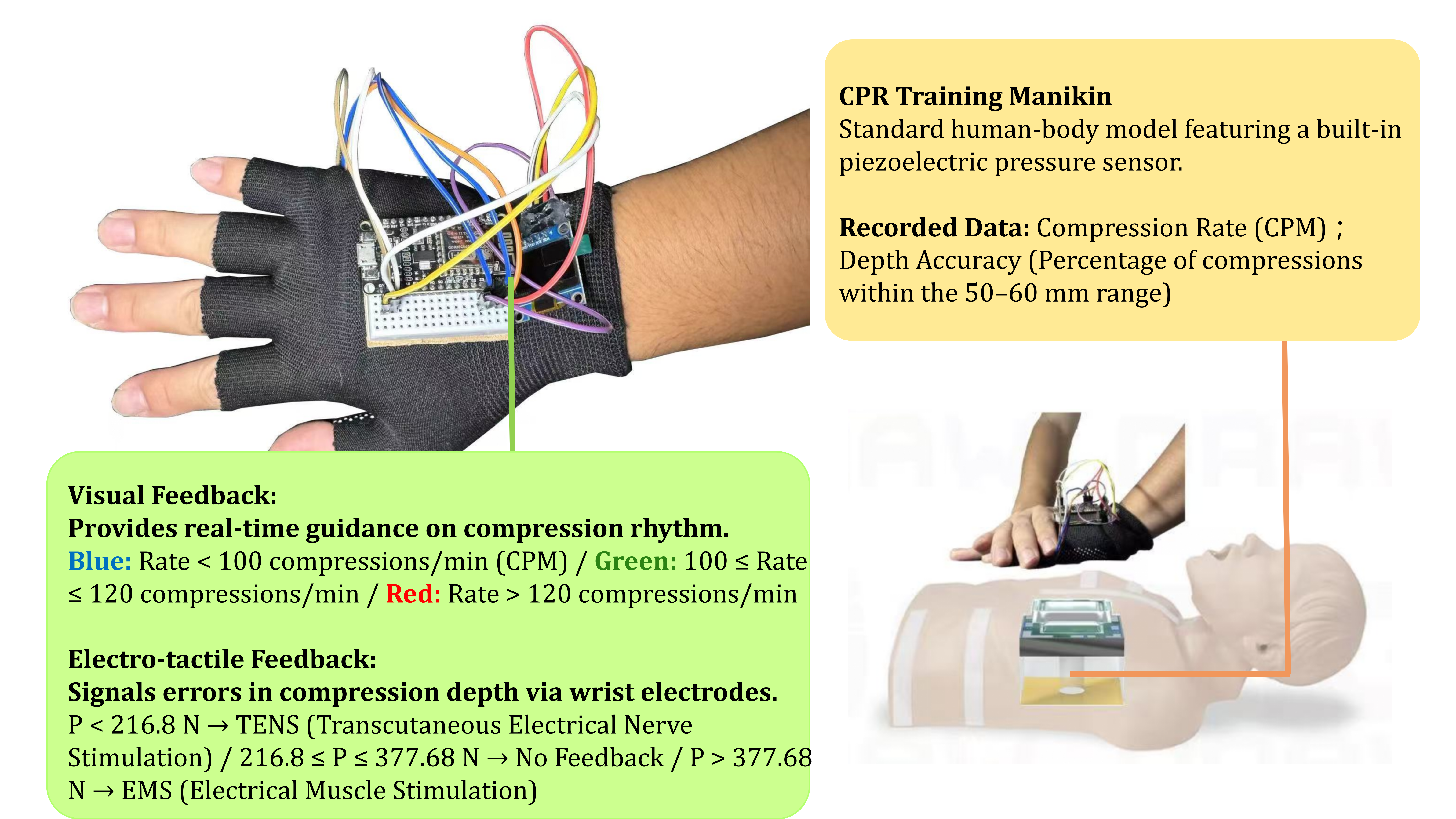}
  \caption{Overview of the TibetCPR hardware. \textit{Upper left:} the wrist-worn unit, comprising a microcontroller and a compact OLED status display housed within a fingerless glove, with two lead wires terminating in hydrogel electrodes that deliver electro-tactile feedback to the user's wrist. \textit{Upper right:} the standard CPR training mannequin with a built-in piezoelectric pressure sensor that records compression rate (CPM) and depth accuracy (the proportion of compressions falling within the 50--60\,mm guideline range). \textit{Lower left:} the two real-time feedback channels---visual rhythm cues (blue / green / red for the three rate zones, displayed on the game-UI screen) and electro-tactile depth signalling (TENS / none / EMS for the three pressure zones, delivered via the wrist electrodes). \textit{Lower right:} a participant performing chest compressions on the mannequin while wearing the wrist-worn unit.}
  \label{fig:example-vertical}
\end{figure}

The hardware system is designed to translate physical chest compressions into in-game events and real-time feedback while remaining robust and low-cost. A high-resolution piezoelectric pressure sensor (measurement range: 0--100~psi; sensitivity: 2.0~mV/V/psi) is embedded within the mannequin torso and continuously tracks compression depth and frequency during gameplay. The system comprises three integrated components: a pressure sensing unit, an electro-tactile stimulation unit, and a visual feedback interface (Fig.~\ref{fig:example-vertical}).

To support real-time haptic feedback, a three-level pressure-to-stimulation mapping is implemented based on international CPR guidelines (Table~\ref{tab:pressure-stimulation}). Electrotactile stimulation is delivered via hydrogel electrodes attached to the user's wrist, enabling both transcutaneous electrical nerve stimulation (TENS) and electrical muscle stimulation (EMS). Stimulation parameters are modulated within controlled ranges (25--50~Hz, 5--20~mA, 200--300~$\mu$s pulse width) to correspond to shallow, adequate, or overly deep compressions.

\begin{table}[t]
\centering
\caption{Pressure-to-stimulation mapping in TibetCPR.}
\label{tab:pressure-stimulation}
\renewcommand{\arraystretch}{1.3}
\begin{tabular}{llllll}
\hline
\textbf{Pressure Range (N)} &
\textbf{Interpretation} &
\textbf{Stimulation Type} &
\textbf{Current (mA)} &
\textbf{Frequency (Hz)} &
\textbf{Duration (s)} \\
\hline
$P < 216.8$ &
\makecell[l]{Inadequate depth\\($<50$ mm)} &
\makecell[l]{Low-frequency\\TENS} &
5--10 &
25 &
10 \\

$216.8 \leq P \leq 377.68$ &
\makecell[l]{Optimal depth\\(50--60 mm)} &
None &
-- &
-- &
-- \\

$P > 377.68$ &
\makecell[l]{Excessive depth\\($>60$ mm)} &
\makecell[l]{High-frequency\\EMS} &
15--20 &
50 &
10 \\
\hline
\end{tabular}
\end{table}

At 25~Hz, low-frequency TENS stimulates superficial A$\beta$ fibers, producing a mild tingling sensation that alerts the user to insufficient compression depth without inducing muscle contraction. At 50~Hz, EMS elicits localized muscle twitches, such as in the pectoral region, signaling overly deep compressions. The electrotactile signal therefore fires only when a compression falls outside the recommended depth range; this boundary-locked, discrete-event design is consistent with prior evidence that electrotactile stimulation supports motor recalibration most effectively when delivered as a discrete error signal~\cite{ravichandran2024electrotactile}, and keeps the wrist channel quiet during compressions that are already in range, preserving the user's attention for the physically demanding action itself.

Compression rhythm is guided through a low-demand visual channel rendered on the game-UI display. Blue indicates compression rates below 100 compressions per minute (CPM), green represents the optimal range of 100--120~CPM, and red warns of rates exceeding 120~CPM. Smooth dynamic transitions between colours reduce perceptual load and allow users to intuitively track rhythm in real time alongside the narrative scene.

The visual and electro-tactile channels are intentionally kept simple and rule-based, with deterministic mappings between sensor readings and feedback outputs that do not vary across users or sessions. This design choice reflects the deployment goal of supporting transparent, autonomously interpretable use across a demographically heterogeneous lay-bystander population, where the semantics of any feedback signal must be inferable through one's own bodily action without external mediation.

%% file: sections/ExperimentalDesign.tex
\subsection{Study Goals and Research Focus}

This study evaluates whether a short, self-guided, narrative-driven CPR training system can support stable acquisition of chest compression skills among CPR-naive lay bystanders in high-altitude, resource-limited regions of Tibet. We focus on interaction-level outcomes that are directly relevant to real-world bystander readiness: whether brief, game-based interaction helps users stabilize compression depth and compression rhythm compared to unassisted practice, and how users from diverse demographic backgrounds experience the system in terms of usability, physical comfort, and perceived independence.

Accordingly, the study investigates (1) changes in chest compression depth accuracy, (2) changes in chest compression rate consistency, and (3) users' subjective experience with the system. These goals reflect the design intent of TibetCPR as a low-burden, self-guided training tool for contexts in which instructor-led CPR training is limited, training opportunities are fragmented, and sustained access to professional supervision cannot be assumed.

\subsection{Participants and Recruitment}

We recruited 40 lay community members in the Tibet Autonomous Region through a combination of offline outreach in Lhasa and Nyingchi---study notices posted in community public spaces and township markets---and online announcements on WeChat, Xiaohongshu, and Bilibili. Recruitment materials described the study as an evaluation of a CPR training game and stated the physical requirements involved in performing chest compressions, allowing prospective participants to self-assess suitability before enrollment.

Participants ranged from 19 to 56 years of age and reflected the broader demographic composition of the region, including both urban and rural residents, occupations spanning farmers, nomadic herders, incense makers, shop owners, tour guides, drivers, teachers, designers, editors, photographers, students, unemployed individuals, and a small number of religious roles (e.g., lama), and educational backgrounds ranging from illiterate or semi-literate through graduate level. None had received prior CPR training or had practical resuscitation experience, and none held a professional background in medicine or emergency response. Demographic details are summarised in Table~\ref{tab:demographics_compact}.

\begin{table}[t]
\centering
\small
\setlength{\tabcolsep}{4pt}
\renewcommand{\arraystretch}{1.1}
\caption{Participant demographics ($N = 40$). Gen.\ = Gender (F = Female, M = Male); U/R = residence (U = Urban, R = Rural); Edu.\ = highest level of education completed.}
\label{tab:demographics_compact}
\begin{tabular}{c c c c l l @{\hspace{1.5em}} c c c c l l}
\hline
\multicolumn{6}{c}{\textbf{Participants 1--20 (Experimental Group)}} &
\multicolumn{6}{c}{\textbf{Participants 21--40 (Control Group)}} \\
\cline{1-6} \cline{7-12}
\textbf{ID} & \textbf{Age} & \textbf{Gen.} & \textbf{U/R} & \textbf{Edu.} & \textbf{Occupation} &
\textbf{ID} & \textbf{Age} & \textbf{Gen.} & \textbf{U/R} & \textbf{Edu.} & \textbf{Occupation} \\
\hline
1  & 27 & F & R & Junior High & Tour guide      & 21 & 24 & M & R & Junior High & Shop owner \\
2  & 33 & M & R & High School & Incense maker   & 22 & 34 & M & R & High School & Driver \\
3  & 34 & F & R & High School & Farmer          & 23 & 33 & M & R & Bachelor    & Lama \\
4  & 26 & M & U & Bachelor    & English teacher & 24 & 33 & F & U & Primary     & Nomadic herder \\
5  & 29 & F & R & Junior High & Farmer          & 25 & 31 & F & U & Bachelor    & Sci.\ communicator \\
6  & 31 & M & U & Semi-lit.   & Unemployed      & 26 & 29 & F & R & High School & Tour guide \\
7  & 35 & F & R & Junior High & Nomadic herder  & 27 & 47 & F & U & Junior High & Farmer \\
8  & 36 & M & R & High School & Unemployed      & 28 & 54 & F & U & High School & Driver \\
9  & 37 & F & U & Primary     & Farmer          & 29 & 34 & M & U & Bachelor    & Math teacher \\
10 & 34 & F & U & Bachelor    & Designer        & 30 & 36 & F & R & High School & Nomadic herder \\
11 & 19 & M & U & High School & Student         & 31 & 45 & M & R & Bachelor    & Math teacher \\
12 & 45 & F & R & Junior High & Incense maker   & 32 & 41 & M & U & High School & Tour guide \\
13 & 47 & M & U & Bachelor    & Editor          & 33 & 42 & F & R & Primary     & Unemployed \\
14 & 50 & F & R & High School & Tour guide      & 34 & 23 & F & U & Junior High & Farmer \\
15 & 51 & M & R & Primary     & Nomadic herder  & 35 & 37 & M & U & High School & Labour \\
16 & 35 & F & U & Junior High & Farmer          & 36 & 19 & M & R & High School & Student \\
17 & 26 & M & R & High School & Shop owner      & 37 & 24 & F & R & Primary     & Nomadic herder \\
18 & 27 & M & R & Master      & Farmer          & 38 & 22 & M & U & Bachelor    & Student \\
19 & 24 & F & U & Bachelor    & Photographer    & 39 & 25 & M & R & Bachelor    & Student \\
20 & 44 & M & R & Bachelor    & Director        & 40 & 19 & M & R & Junior High & Student \\
\hline
\end{tabular}
\end{table}

Inclusion criteria were: at least 18 years of age; physically capable of performing short-duration chest compressions; no prior CPR training; and no professional background in medicine or emergency response. Exclusion criteria included pre-existing cardiovascular disease, severe musculoskeletal injury, or other conditions that could pose risk during chest compressions, and cognitive or language conditions that could interfere with comprehension of task instructions.

The study was reviewed and approved by the Institutional Review Board (IRB) of [Authors' Institution]. Before participation, all individuals received a written informed consent form describing the study's purpose, procedures, data collection, the physical demands and short-term fatigue of chest compressions, and their right to withdraw at any time without penalty. All data were anonymised prior to analysis, and each participant received 50~RMB upon completion. Participants were randomly assigned to an experimental group ($n = 20$, P1--P20) or a control group ($n = 20$, P21--P40); baseline equivalence in compression performance is assessed using T1 measurements and addressed analytically in the Results.

\subsection{Procedure}

The study employed a randomized controlled design comprising three phases: pre-test (T1), intervention (T2), and post-test (T3). The procedure was structured to approximate how CPR training would realistically occur in high-altitude, resource-limited regions, where training is typically self-initiated, brief, and conducted without professional supervision; the design therefore prioritized ecological plausibility for short, autonomous training encounters under physical and infrastructural constraints.

Upon arrival, participants were introduced to the study environment and equipment. They were informed that their CPR performance and interaction behaviors would be recorded automatically by the system for research purposes, and that experimental-group participants would additionally take part in a short interview after the post-test. No additional instructional guidance was provided beyond basic system orientation, reflecting the absence of instructor mediation in many real-world training situations in the region.

\textbf{T1: Pre-test.}
All participants performed a one-minute unassisted chest-compression session on a standard CPR training mannequin. The mannequin was instrumented with an embedded pressure sensor that recorded compression depth and frequency but provided no other cues; no visual, tactile, or auditory feedback was delivered during T1. This baseline phase was designed to capture each participant's initial ability to regulate compression depth and rhythm in the absence of external prompts, while minimizing physical fatigue prior to the main intervention. Compression depth and compression rate were recorded automatically by the system.

\textbf{T2: Intervention.}
After a brief orientation to the TibetCPR system, participants in the experimental group completed a 10-minute training session in which the system delivered real-time multimodal feedback: depth-driven electro-tactile stimulation paired with rhythm-driven visual cues. Participants in the control group performed 10 minutes of unassisted practice on the same mannequin without any system feedback, reflecting unguided practice scenarios in which individuals attempt to improve CPR technique through repetition alone. The two groups were matched in training duration, task, and equipment; only the presence of real-time feedback differed. The intervention duration was intentionally limited to reflect realistic opportunities for opportunistic practice in public or educational settings and to avoid excessive physical strain from sustained chest compressions at high altitude. Compression data were recorded continuously for both groups.

\textbf{T3: Post-test and Reflection.}
All participants then completed a second one-minute unassisted chest-compression session identical to T1, allowing us to assess compression performance once system support was removed. Participants in the experimental group subsequently completed the System Usability Scale (SUS) and a semi-structured interview about their experience with the training process and the feedback mechanisms.

This study was intended as a formative evaluation of interaction-level training effectiveness under realistic deployment constraints. Given the emphasis on short, self-guided use, physical-exertion considerations, and the exploratory goal of examining whether brief training encounters can support motor-skill stabilization, a moderate sample size and a limited intervention duration were adopted.

\subsection{Measures and Data Collection}

\textbf{Compression Performance.}
Compression depth and rate were recorded automatically by the embedded pressure sensor throughout each CPR session, providing observer-independent metrics for both groups. Compression rate was computed as compressions per minute (CPM); rhythm deviation was the absolute difference from the guideline-recommended target of 110~CPM, with smaller deviations indicating more stable rhythm \cite{meaney2013cardiopulmonary}. Compression depth was evaluated against the 50--60~mm guideline range \cite{stiell2012role}; the proportion of compressions within this range was used as the depth-control measure.

\textbf{Usability and User Experience.}
System usability was assessed using the System Usability Scale (SUS), a standardised 10-item questionnaire rated on a five-point Likert scale (Table~\ref{tab:sus-content}). Responses to negatively worded items (Items~2, 4, 6, 8, 10) were reverse-coded so that higher values on every item correspond to more favourable usability; the overall SUS score was then computed via the standard scoring procedure on a $0$--$100$ scale and interpreted using Bangor et~al.'s adjective rating ranges. Qualitative data were collected through semi-structured interviews of approximately 20--25 minutes; with participant consent, interviews were audio-recorded and subsequently transcribed for analysis.

\textbf{Artifact and Data Availability.}
To support reproducibility and further deployment in resource-constrained settings, the wearable unit's hardware design (schematics and bill of materials), the microcontroller firmware, the game-UI source code, and the de-identified per-compression interaction data will be released upon publication.

\begin{table}[t]
\centering
\caption{SUS content.}
\label{tab:sus-content}
\renewcommand{\arraystretch}{1.3}
\begin{tabular}{clc}
\hline
\textbf{Item No.} & \textbf{Statement} & \textbf{Score} \\
\hline
1 &
\makecell[l]{I think I would be willing to play this game frequently\\
to practice CPR.} &
\\

2 &
I find this game too complex. &
\\

3 &
I think the operation of this game is very simple. &
\\

4 &
\makecell[l]{I need help from others to understand how to play\\
this game.} &
\\

5 &
\makecell[l]{I find that various functions in the game (such as\\
tactile feedback, visual prompts) work well together.} &
\\

6 &
I find the rules and feedback in the game inconsistent. &
\\

7 &
\makecell[l]{I think most people can quickly learn how to play\\
this game.} &
\\

8 &
I find the operation of the game awkward. &
\\

9 &
I am confident in playing this game independently. &
\\

10 &
\makecell[l]{Before starting to play, I need to learn a lot of\\
operating instructions.} &
\\
\hline
\end{tabular}
\end{table}

\subsection{Data Analysis}

\subsubsection{Quantitative Analysis}

The study followed a mixed-design (between-subject group $\times$ within-subject time) randomised controlled layout, with two co-primary outcomes: compression rate deviation from the 110~CPM target (rhythm) and compression depth accuracy in the 50--60~mm range (depth). Analyses were conducted in Python with deterministic random seeds. Throughout, we used Shapiro--Wilk and Levene's tests as diagnostic checks; where Levene's test indicated unequal variances, we used Welch's $t$-test in place of Student's $t$. Random-allocation balance at T1 was additionally cross-checked with Mann--Whitney $U$ as a non-parametric robustness test.

The primary T1$\rightarrow$T3 contrast was modelled with analysis of covariance (ANCOVA), regressing T3 on Group with T1 as a covariate. We tested the homogeneity-of-regression-slopes assumption via the Group$\times$T1 interaction; where it was violated, we treated the ANCOVA $F$ and partial $\eta^2$ as descriptive and rested inference on triangulating estimates: within-group paired $t$-tests (Cohen's $d_z$), a between-group independent $t$-test on change scores $\Delta = T3 - T1$ (Cohen's $d$), and percentile-bootstrap 95\% confidence intervals (10{,}000 resamples).

Within-session trajectories across the 10-minute T2 phase were analysed with a linear mixed-effects model regressing the per-minute outcome on Group, centred Minute, and their interaction, with a random intercept per participant. The Group$\times$Minute interaction provided the principal test of progressive divergence between conditions; we preferred this specification over repeated-measures ANOVA because it does not require sphericity and yields directly interpretable per-minute slope estimates. We additionally contrasted minute-1 with minute-10 using paired $t$-tests within group and independent $t$-tests between groups.

For every contrast we report the test statistic with degrees of freedom, exact $p$-value, an effect size (Cohen's $d$, $d_z$, or partial $\eta^2$ as appropriate), and a 95\% confidence interval (parametric where assumptions held; percentile-bootstrap otherwise). All tests used $\alpha = 0.05$; the two co-primary outcomes were specified \textit{a priori} as separate hypotheses, and Holm--Bonferroni adjustment did not alter the headline conclusions. SUS responses were summarised descriptively with item-level means and standard deviations and the overall 0--100 score.

\subsubsection{Qualitative Analysis}

Semi-structured interviews were audio-recorded with participant consent and transcribed verbatim within 12 hours of collection. Transcripts were analysed using thematic analysis following an iterative, inductive coding process \cite{braun2012thematic}. Two researchers independently coded the qualitative data using an initial codebook reflecting participants' perceptions of feedback clarity, physical comfort, learning confidence, and autonomy during training. Inter-rater reliability was quantified using Cohen's $\kappa$, with a result of $\kappa = 0.84$, indicating almost perfect agreement between coders. Across two iterative coding cycles, the researchers discussed and resolved definitional ambiguities and any coding disagreements until consensus was reached on all codes. The agreed-upon codes were then consolidated into the higher-level themes reported in Section~5.3 through repeated reference to the original transcripts.

%% file: sections/Results.tex
\subsection{Mastering the Beat: Compression Rhythm Stabilization}

We first examined whether interaction with TibetCPR changed how stably participants regulated chest compression rhythm, operationalised as the absolute deviation of compressions per minute from the guideline-recommended 110~CPM target. Analyses focused on (i)~pre-to-post change between T1 and T3, and (ii)~the within-session trajectory across the 10-minute intervention phase (T2), consistent with the system's design intent of supporting short, self-guided practice in instructor-limited settings. As a first step, we verified that random allocation produced equivalent baseline performance: T1 rate deviation did not differ between the experimental ($M = 30.65$, $SD = 15.06$) and control ($M = 32.65$, $SD = 16.02$) groups, $t(38) = -0.41$, $p = .69$, Cohen's $d = -0.13$, with a 95\% bootstrap confidence interval for the between-group difference of $[-11.20, 7.40]$. Both groups began the study with comparable difficulty maintaining a stable compression pace.

\paragraph{Post-training rhythm changes (T1 to T3).}

\begin{figure}[htbp]
  \centering
  \begin{subfigure}[t]{0.485\linewidth}
    \centering
    \includegraphics[width=\linewidth]{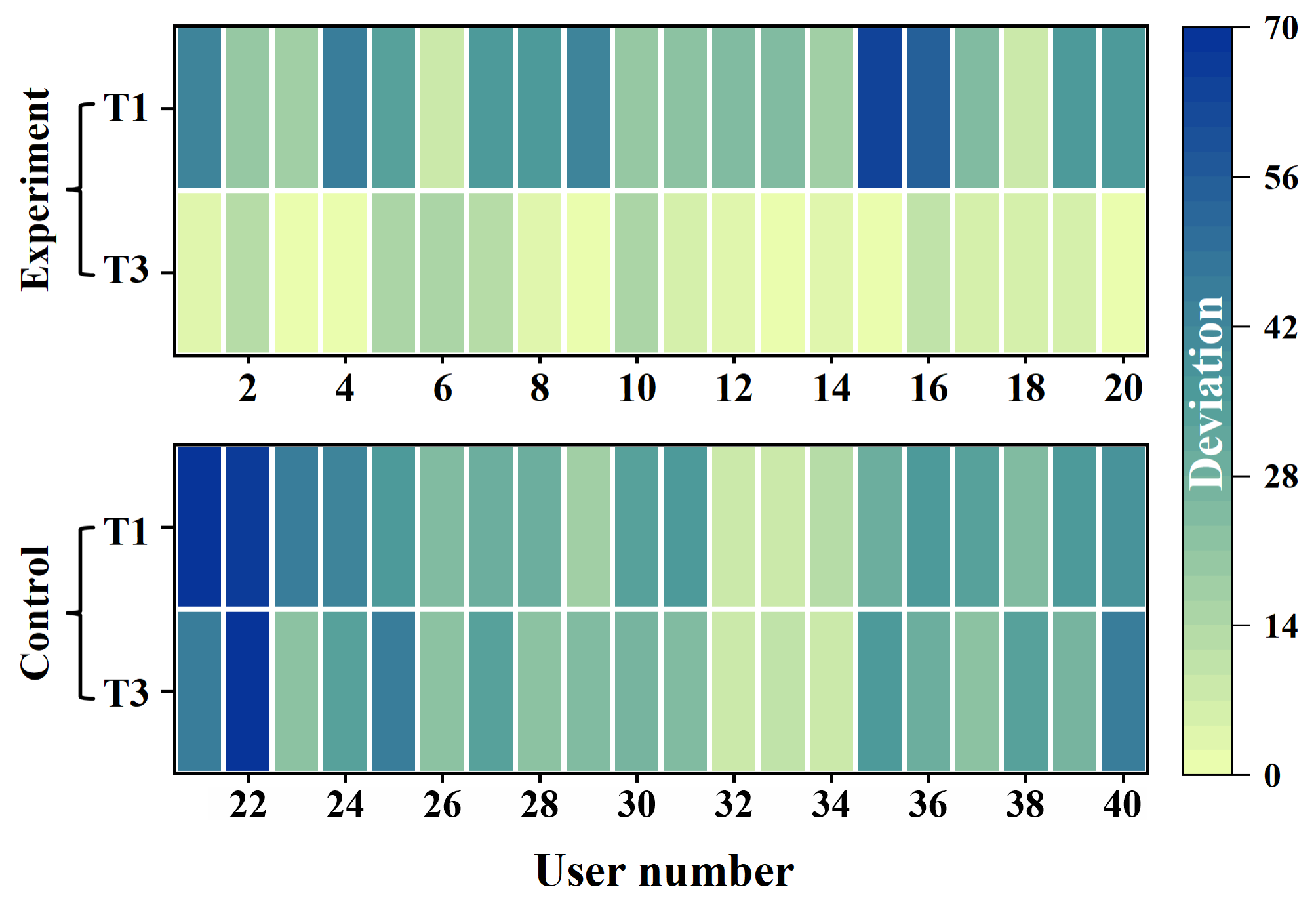}
    %\caption{First image caption}
    \label{fig:example-a}
  \end{subfigure}
  \hfill
  \begin{subfigure}[t]{0.495\linewidth}
    \centering
    \includegraphics[width=\linewidth]{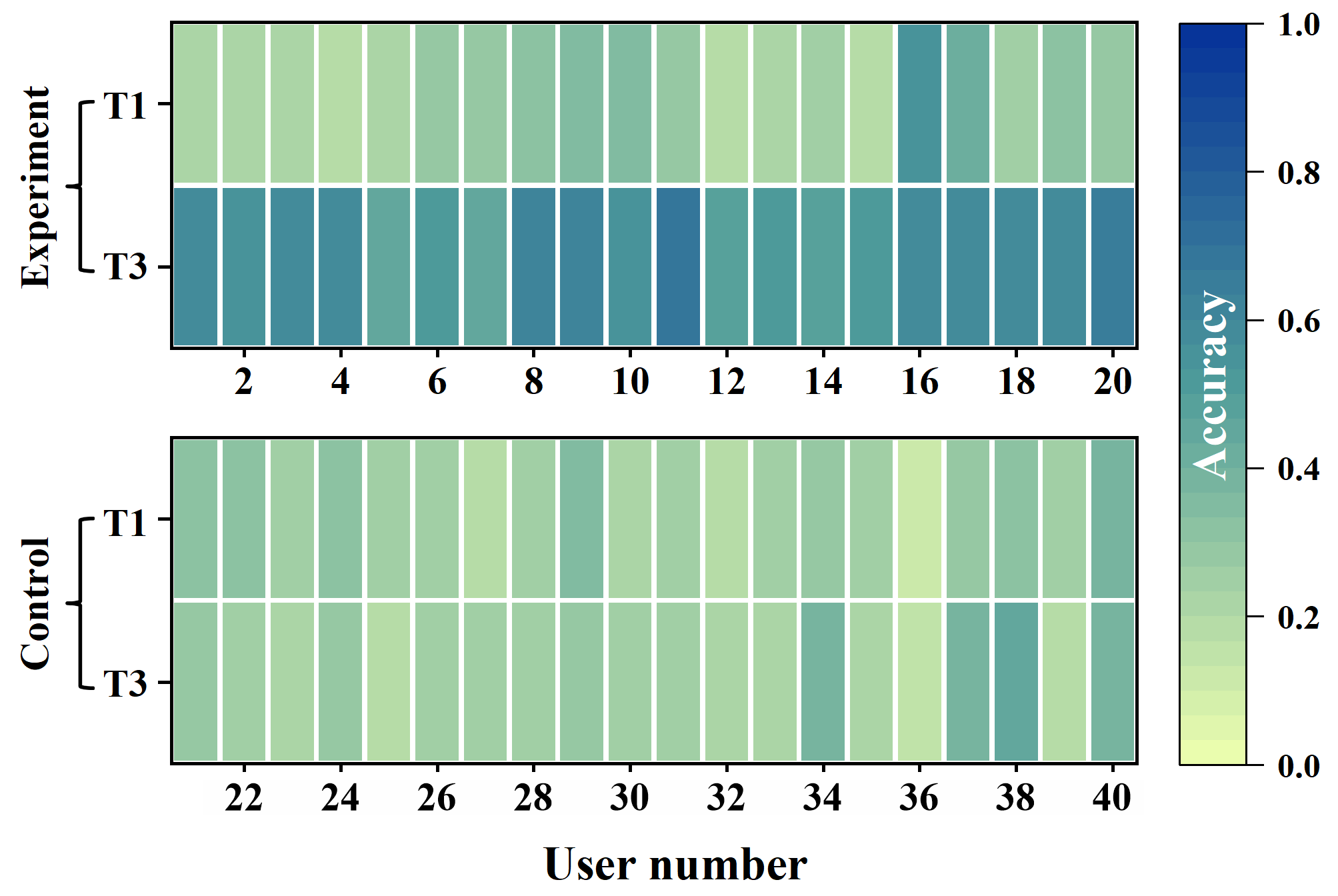}
    %\caption{Second image caption}
    \label{fig:example-b}
  \end{subfigure}
  \caption{Individual-level performance at the pre-intervention (T1) and post-intervention (T3) stages for all 40 participants, with the experimental group shown in the upper sub-row and the control group in the lower sub-row of each panel: (a)~heatmap of compression rate deviation from the target of 110~CPM (rhythm); (b)~heatmap of compression depth accuracy (proportion of compressions within the recommended 50--60\,mm range).}
  \label{fig:rate-individual}
\end{figure}

\begin{figure}[htbp]
  \centering
  \begin{subfigure}[t]{0.485\linewidth}
    \centering
    \includegraphics[width=\linewidth]{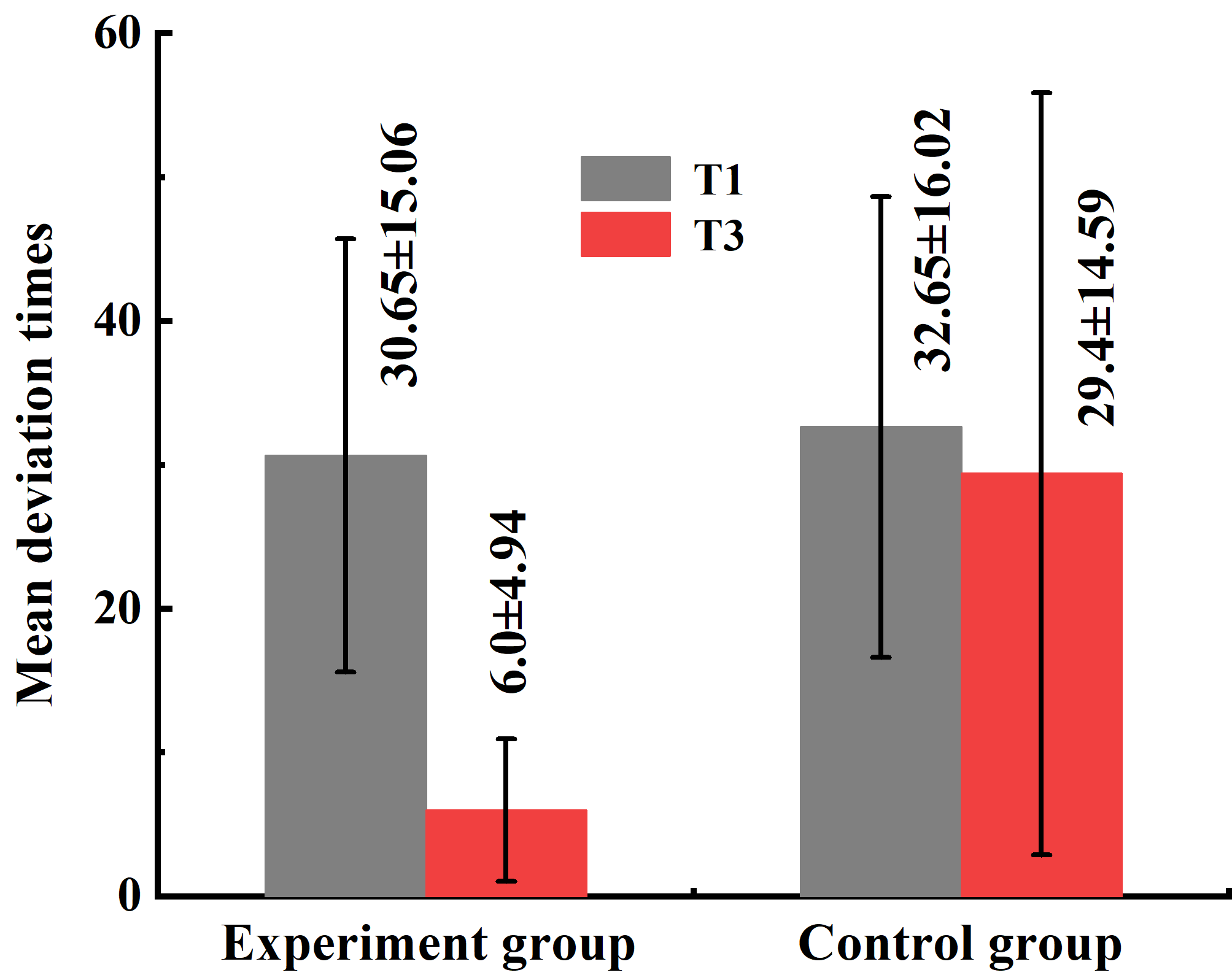}
    \label{fig:rate-mean-a}
  \end{subfigure}
  \hfill
  \begin{subfigure}[t]{0.495\linewidth}
    \centering
    \includegraphics[width=\linewidth]{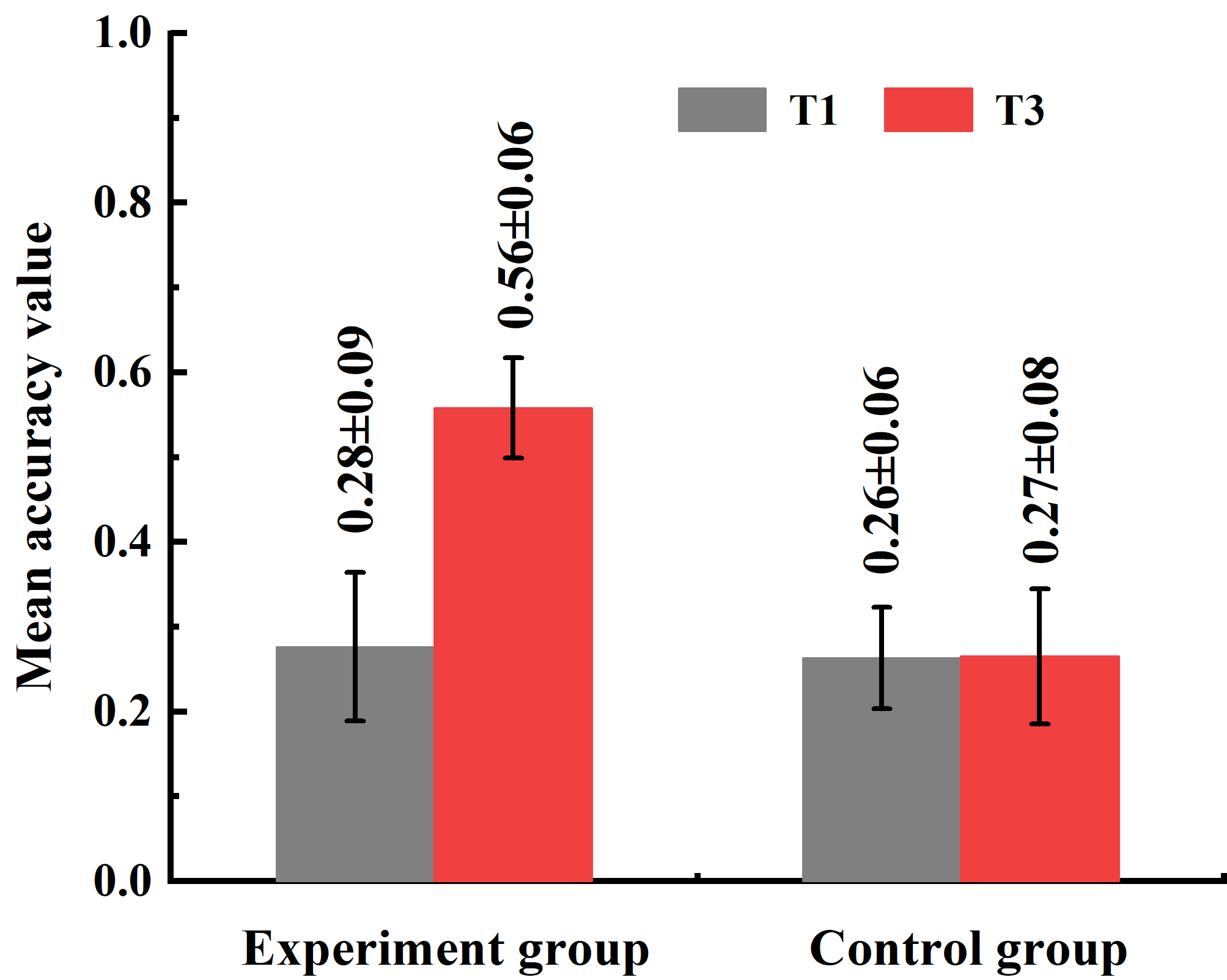}
    \label{fig:rate-mean-b}
  \end{subfigure}
  \caption{Group-level summary of pre-intervention (T1) to post-intervention (T3) performance for the experimental and control groups: (a)~mean compression rate deviation from the target of 110~CPM (rhythm); (b)~mean compression depth accuracy (proportion of compressions within the recommended 50--60\,mm range). Bars show T1 and T3 means with standard-deviation error bars; numerical labels denote mean~$\pm$~standard deviation.}
  \label{fig:rate-mean}
\end{figure}

Figure~\ref{fig:rate-individual}(a) presents individual-level rate deviation for both groups at T1 and T3, and Figure~\ref{fig:rate-mean}(a) summarises the corresponding group-level means. After the 10-minute training session, experimental participants' rate deviation fell from $M = 30.65$ ($SD = 15.06$) at T1 to $M = 6.00$ ($SD = 4.94$) at T3---a within-group reduction of $24.65$ deviations on average ($t(19) = -6.35$, $p < .001$, Cohen's $d_z = -1.42$, 95\% bootstrap CI for the change $[-32.20, -17.30]$). The control group, who completed identical practice without system feedback, showed only a small and non-significant shift, from $M = 32.65$ ($SD = 16.02$) at T1 to $M = 29.40$ ($SD = 14.59$) at T3 ($t(19) = -1.48$, $p = .156$, $d_z = -0.33$).

A direct between-group comparison of the change scores ($\Delta = T3 - T1$) confirmed that the experimental gain substantially exceeded any change observed in control, $t(38) = -4.80$, $p < .001$, Cohen's $d = -1.52$, 95\% bootstrap CI for $\Delta_{\mathrm{Exp}} - \Delta_{\mathrm{Ctrl}}$ $[-29.95, -13.00]$. An ANCOVA controlling for T1 baseline mirrored this conclusion at post-test, $F(1, 37) = 54.64$, $p < .001$, partial $\eta^2 = .596$. Within the experimental group, improvements scaled with initial baseline performance: participants who began with larger initial deviation showed the largest reductions, suggesting that the system was particularly effective at stabilising users starting from less consistent baselines. Crucially, these post-training differences emerged after the system was switched off: participants' improved pacing transferred to an unscaffolded one-minute post-test, rather than depending on real-time feedback being present at the moment of measurement.

\paragraph{Within-session rhythm dynamics during the intervention (T2).}

\begin{figure}[htbp]
  \centering
  \includegraphics[width=\linewidth]{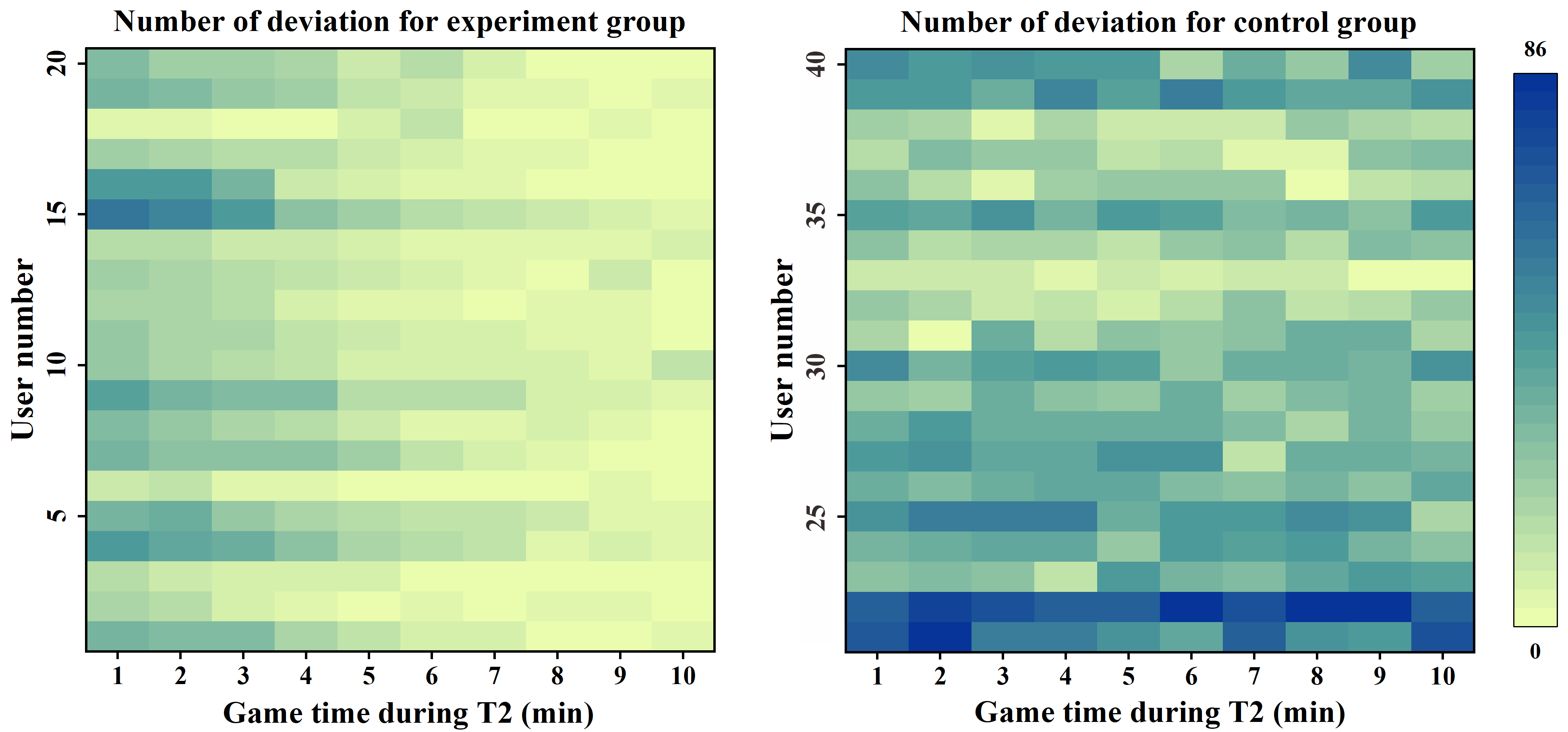}
  \caption{Per-minute compression rate deviation from the target of 110~CPM during the 10-minute intervention phase (T2), shown as a heatmap with each row corresponding to one participant and each column to one minute of training: (a, left)~experimental group (P1--P20); (b, right)~control group (P21--P40). Lighter shades indicate lower deviation.}
  \label{fig:rate-heatmap}
\end{figure}

\begin{figure}[htbp]
  \centering
  \begin{subfigure}[t]{0.495\linewidth}
    \centering
    \includegraphics[width=\linewidth]{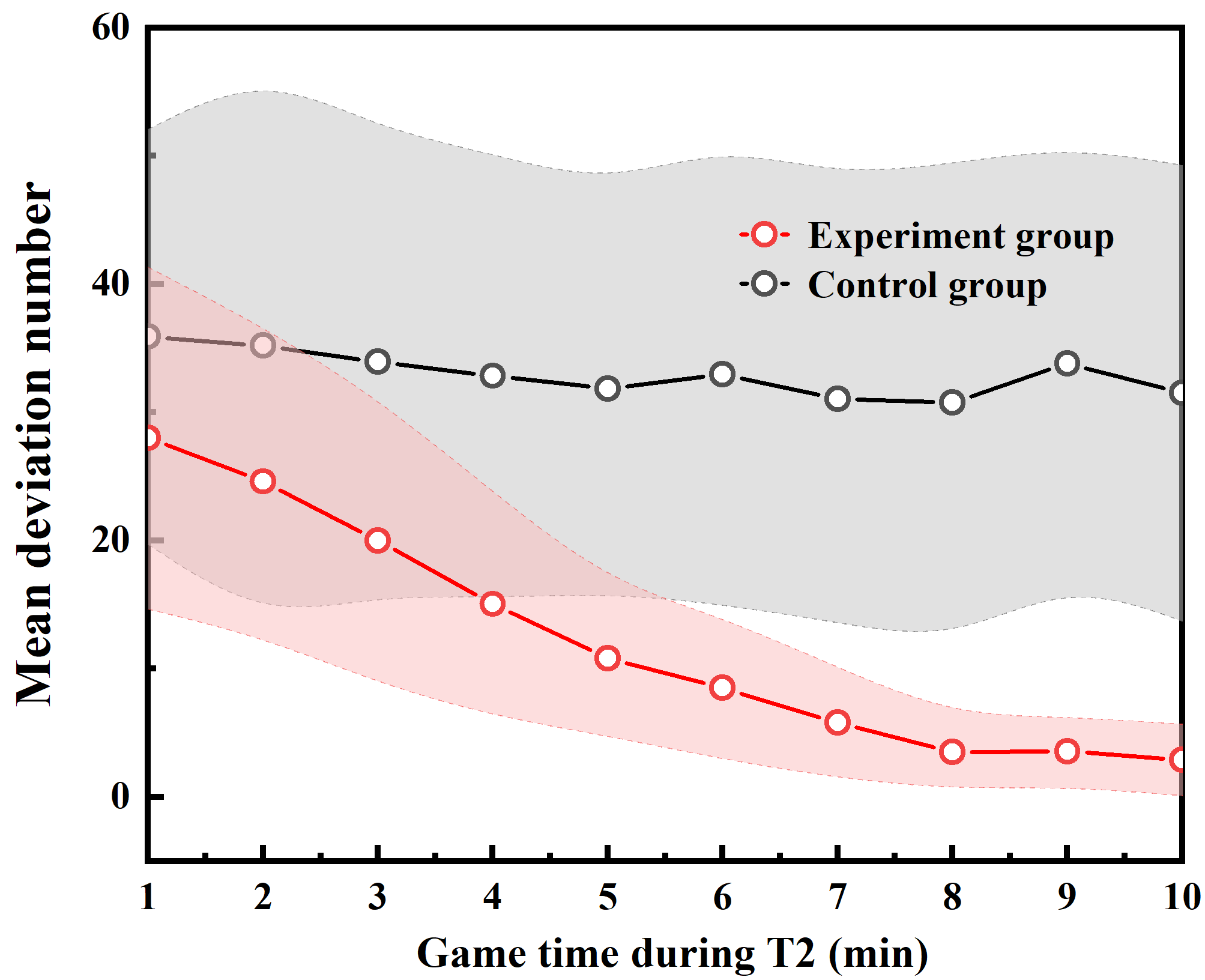}
    \label{fig:rate-trajectory-a}
  \end{subfigure}
  \hfill
  \begin{subfigure}[t]{0.495\linewidth}
    \centering
    \includegraphics[width=\linewidth]{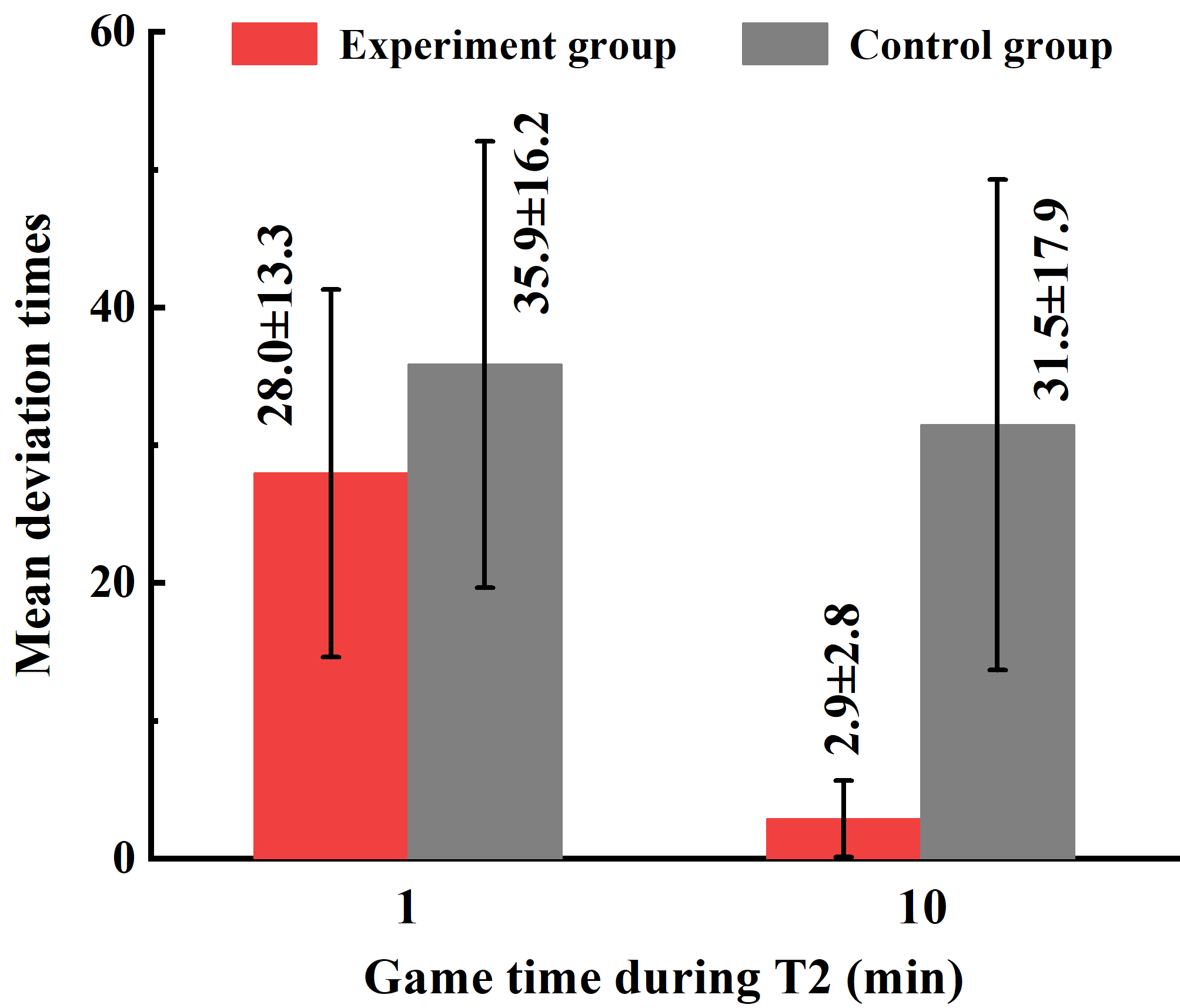}
    \label{fig:rate-trajectory-b}
  \end{subfigure}
  \caption{Compression rate deviation from the target of 110~CPM during the intervention phase (T2) for the experimental and control groups: (a)~mean trajectory across the 10-minute training session, with shaded bands showing $\pm 1$ standard deviation; (b)~mean deviation at the first and last minute of training (mean $\pm$ standard deviation).}
  \label{fig:rate-trajectory}
\end{figure}

To examine how rhythm regulation evolved while participants were actively using the system, we analysed minute-by-minute deviation across the 10-minute T2 phase. Figure~\ref{fig:rate-heatmap} visualises individual participants' deviation values over time, with the experimental group on the left and the control group on the right. In the experimental group, deviation values became progressively lower and more concentrated as training continued, and most participants converged to near-zero deviation in the final minutes of T2. The control group showed no comparable temporal pattern: deviation values remained high and dispersed throughout the intervention phase. We formalised this observation with a linear mixed-effects model regressing per-minute deviation on Group, centred Minute, and their interaction, with a random intercept per participant. The Group$\times$Minute interaction was highly significant, $\beta = -2.52$ deviations per minute, $p < .001$: the experimental group's deviation decreased approximately 2.5 units per minute faster than the control group's, capturing the gradual stabilisation visible in the trajectory plotted in Figure~\ref{fig:rate-trajectory}(a).

A direct contrast between the first and last minute of T2 (Fig.~\ref{fig:rate-trajectory}(b)) made the boundary effect clear. At minute~1, experimental ($M = 28.00$, $SD = 13.33$) and control ($M = 35.90$, $SD = 16.19$) participants did not differ significantly, $t(38) = -1.68$, $p = .100$, Cohen's $d = -0.53$, with a 95\% bootstrap CI for the difference of $[-16.90, 0.95]$. By minute~10, the experimental group had reduced deviation to $M = 2.90$ ($SD = 2.79$) while the control group remained at $M = 31.50$ ($SD = 17.79$); the gap was both large and statistically clear, $t = -7.10$ (Welch's), $p < .001$, $d = -2.25$, 95\% bootstrap CI $[-36.55, -21.35]$. Within the experimental group, the change from minute~1 to minute~10 was substantial ($t(19) = -8.48$, $p < .001$, paired $d_z = -1.90$); within the control group, the same contrast did not reach significance ($t(19) = -1.75$, $p = .096$, $d_z = -0.39$), suggesting that comparable practice without feedback was insufficient to drive rhythm stabilisation in this short timeframe.

Together, these results indicate that compression rhythm stabilisation in the experimental group emerged progressively over the course of interaction and transferred to an unscaffolded post-test, while comparable practice without feedback did not yield a similar trajectory. This temporal pattern aligns with the system's design intent of supporting rhythm calibration through interaction itself rather than through continuous external prescription, and is consistent with the broader hypothesis that short, self-guided CPR training can support stable motor performance under conditions where instructor mediation is unavailable.

\subsection{Compression Depth Accuracy and Usability}

We next examined whether interaction with TibetCPR changed how accurately participants regulated compression depth, operationalised as the proportion of compressions falling within the guideline-recommended 50--60~mm range. As with rhythm, analyses focused on (i)~pre-to-post change between T1 and T3, and (ii)~the within-session trajectory across T2. Random allocation again produced equivalent baseline performance: T1 depth accuracy did not differ between the experimental ($M = 0.277$, $SD = 0.088$) and control ($M = 0.264$, $SD = 0.060$) groups, $t(38) = 0.55$, $p = .59$, Cohen's $d = 0.17$, with a 95\% bootstrap CI for the between-group difference of $[-0.030, 0.061]$. Both groups began the study with low and similarly variable depth control.

\paragraph{Post-training depth accuracy changes (T1 to T3).}

Figure~\ref{fig:rate-individual}(b) presents individual-level depth accuracy at T1 and T3 for all 40 participants, and Figure~\ref{fig:rate-mean}(b) summarises the corresponding group-level means. Following the 10-minute training session, experimental participants' accuracy rose from $M = 0.277$ ($SD = 0.088$) at T1 to $M = 0.558$ ($SD = 0.059$) at T3---a within-group gain of $0.282$ on average ($t(19) = 13.96$, $p < .001$, Cohen's $d_z = 3.12$, 95\% bootstrap CI for the change $[0.240, 0.317]$). The control group, who completed identical practice without system feedback, showed essentially no change, from $M = 0.264$ ($SD = 0.060$) at T1 to $M = 0.265$ ($SD = 0.079$) at T3 ($t(19) = 0.13$, $p = .896$, $d_z = 0.03$).

A direct between-group comparison of the change scores ($\Delta = T3 - T1$) confirmed that the experimental gain was substantially larger than the control change, $t(38) = 11.56$, $p < .001$, Cohen's $d = 3.66$, 95\% bootstrap CI for $\Delta_{\mathrm{Exp}} - \Delta_{\mathrm{Ctrl}}$ $[0.231, 0.323]$. An ANCOVA controlling for T1 baseline mirrored this conclusion at post-test, $F(1, 37) = 203.17$, $p < .001$, partial $\eta^2 = .846$. As with rhythm, improvements within the experimental group scaled with initial baseline performance: participants who began with lower initial accuracy showed the largest gains, again consistent with the system being most effective for users starting from less consistent baselines. Crucially, these post-training improvements emerged after the system was switched off: the experimental group's depth-accuracy gains transferred to an unscaffolded one-minute post-test rather than depending on real-time feedback being present.

\paragraph{Depth regulation during the intervention (T2).}

\begin{figure}[htbp]
  \centering
  \includegraphics[width=\linewidth]{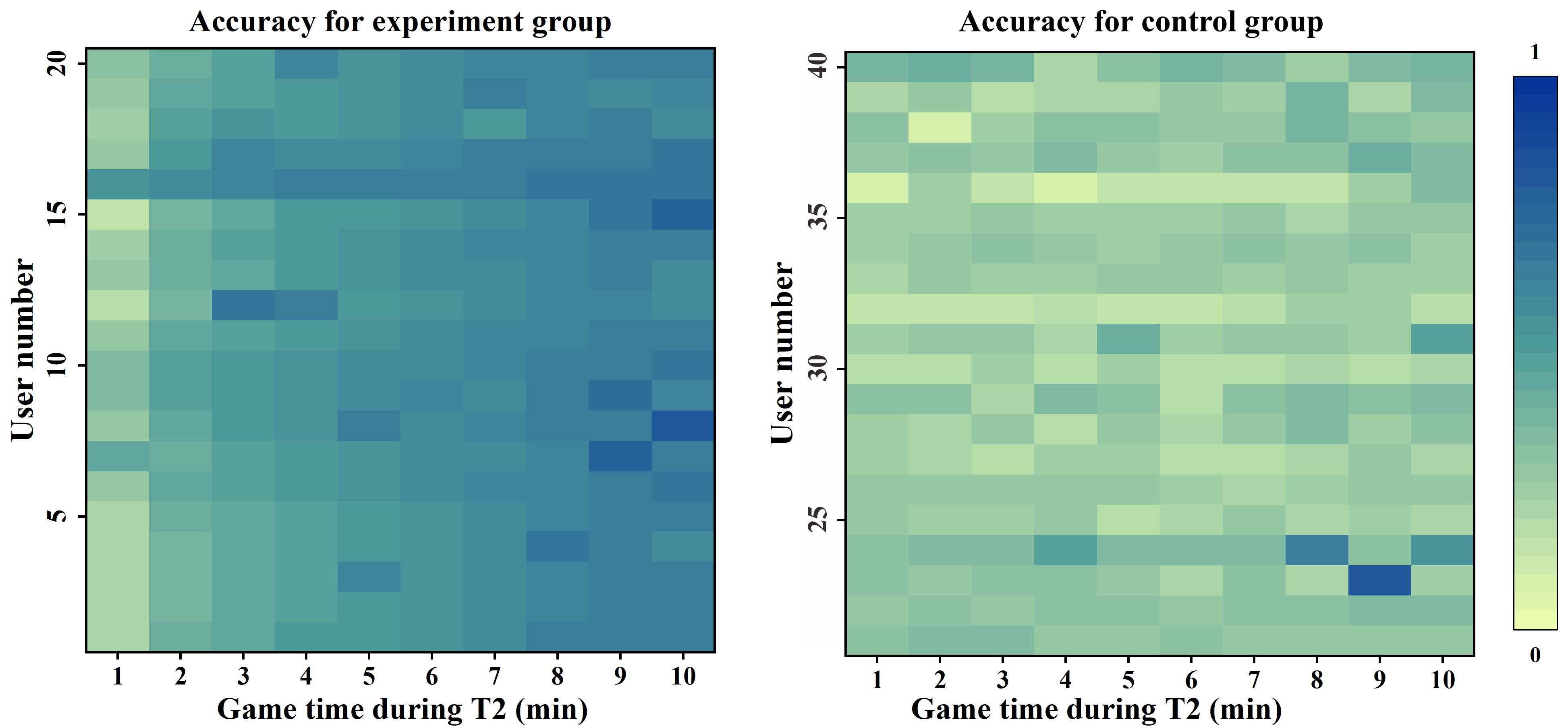}
  \caption{Per-minute compression depth accuracy during the 10-minute intervention phase (T2), shown as a heatmap with each row corresponding to one participant and each column to one minute of training: (a, left)~experimental group (P1--P20); (b, right)~control group (P21--P40). Darker shades indicate higher accuracy.}
  \label{fig:depth-heatmap}
\end{figure}

\begin{figure}[htbp]
  \centering
  \begin{subfigure}[t]{0.495\linewidth}
    \centering
    \includegraphics[width=\linewidth]{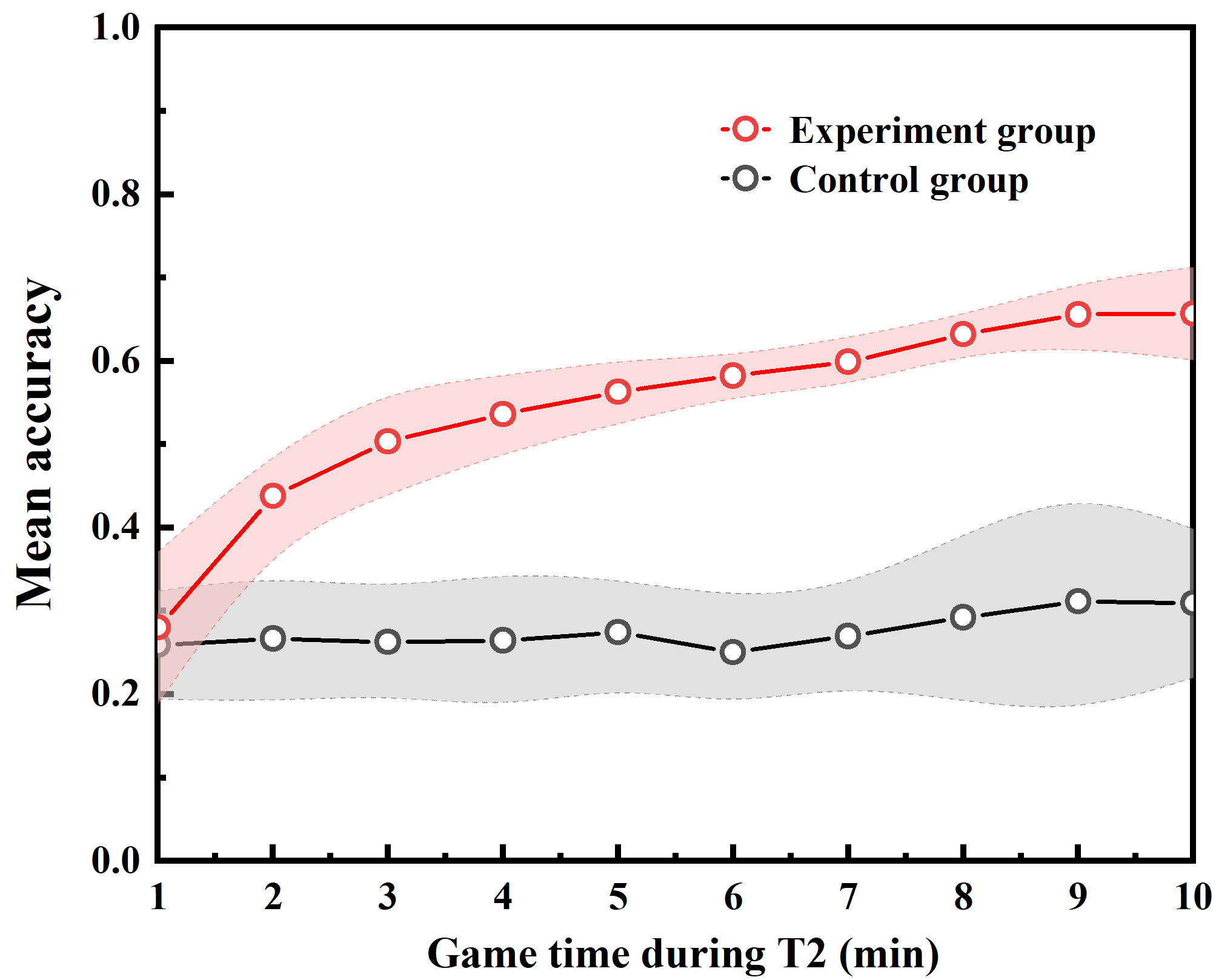}
    \label{fig:depth-trajectory-a}
  \end{subfigure}
  \hfill
  \begin{subfigure}[t]{0.495\linewidth}
    \centering
    \includegraphics[width=\linewidth]{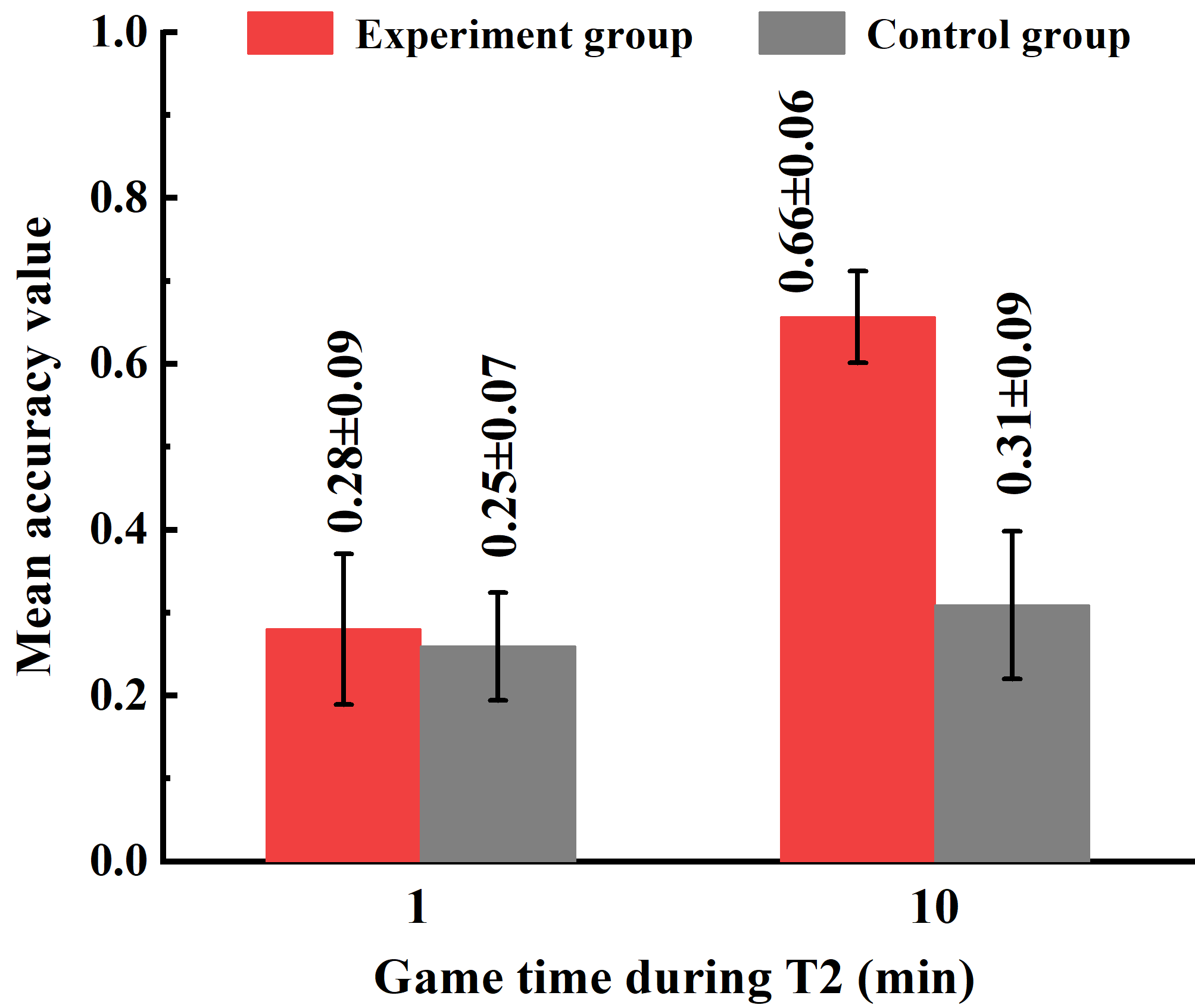}
    \label{fig:depth-trajectory-b}
  \end{subfigure}
  \caption{Compression depth accuracy during the intervention phase (T2) for the experimental and control groups: (a)~mean accuracy across the 10-minute training session, with shaded bands showing $\pm 1$ standard deviation; (b)~mean accuracy at the first and last minute of training (mean $\pm$ standard deviation).}
  \label{fig:depth-trajectory}
\end{figure}

To examine how depth control evolved while participants were actively using the system, we analysed minute-by-minute accuracy across the 10-minute T2 phase. Figure~\ref{fig:depth-heatmap} visualises individual participants' accuracy values over time, with the experimental group on the left and the control group on the right. In the experimental group, accuracy rose progressively across the session, with most participants stabilising at substantially higher accuracy by the later minutes of T2. The control group showed a comparatively flat pattern, with accuracy values remaining low and dispersed throughout the intervention phase. We formalised this observation with a linear mixed-effects model regressing per-minute accuracy on Group, centred Minute, and their interaction, with a random intercept per participant. The Group$\times$Minute interaction was highly significant, $\beta = +0.030$ per minute, $p < .001$: the experimental group's accuracy rose roughly three percentage points per minute faster than the control group's, capturing the gradual improvement visible in the trajectory plotted in Figure~\ref{fig:depth-trajectory}(a).

A direct contrast between the first and last minute of T2 (Fig.~\ref{fig:depth-trajectory}(b)) clarified the boundary effect. At minute~1, experimental ($M = 0.280$, $SD = 0.091$) and control ($M = 0.259$, $SD = 0.065$) participants did not differ significantly, $t(38) = 0.83$, $p = .411$, Cohen's $d = 0.26$, with a 95\% bootstrap CI for the difference of $[-0.025, 0.071]$. By minute~10, experimental accuracy had risen to $M = 0.657$ ($SD = 0.056$) while control accuracy reached $M = 0.309$ ($SD = 0.089$); the gap was both large and statistically clear, $t(38) = 14.80$, $p < .001$, $d = 4.68$, 95\% bootstrap CI $[0.302, 0.392]$. Within the experimental group, the change from minute~1 to minute~10 was striking ($t(19) = 16.23$, $p < .001$, paired $d_z = 3.63$). The control group also showed a small but significantly increasing trend ($t(19) = 2.38$, $p = .028$, paired $d_z = 0.53$), suggesting a modest practice effect from repeated execution; this within-session bump, however, was approximately seven times smaller than the experimental gain and---as the T1$\rightarrow$T3 contrast above showed---did not carry through to the unscaffolded post-test.

Together, these results indicate that compression depth accuracy in the experimental group rose progressively over the course of interaction and transferred to the unscaffolded post-test, while comparable practice without feedback yielded only a small within-session bump that did not carry over to T3. As with rhythm, this temporal pattern is consistent with the system's design intent of supporting motor calibration through interaction itself rather than through continuous external prescription.

\paragraph{System usability and perceived independence.}

\begin{figure}[htbp]
  \centering
  \includegraphics[width=\linewidth]{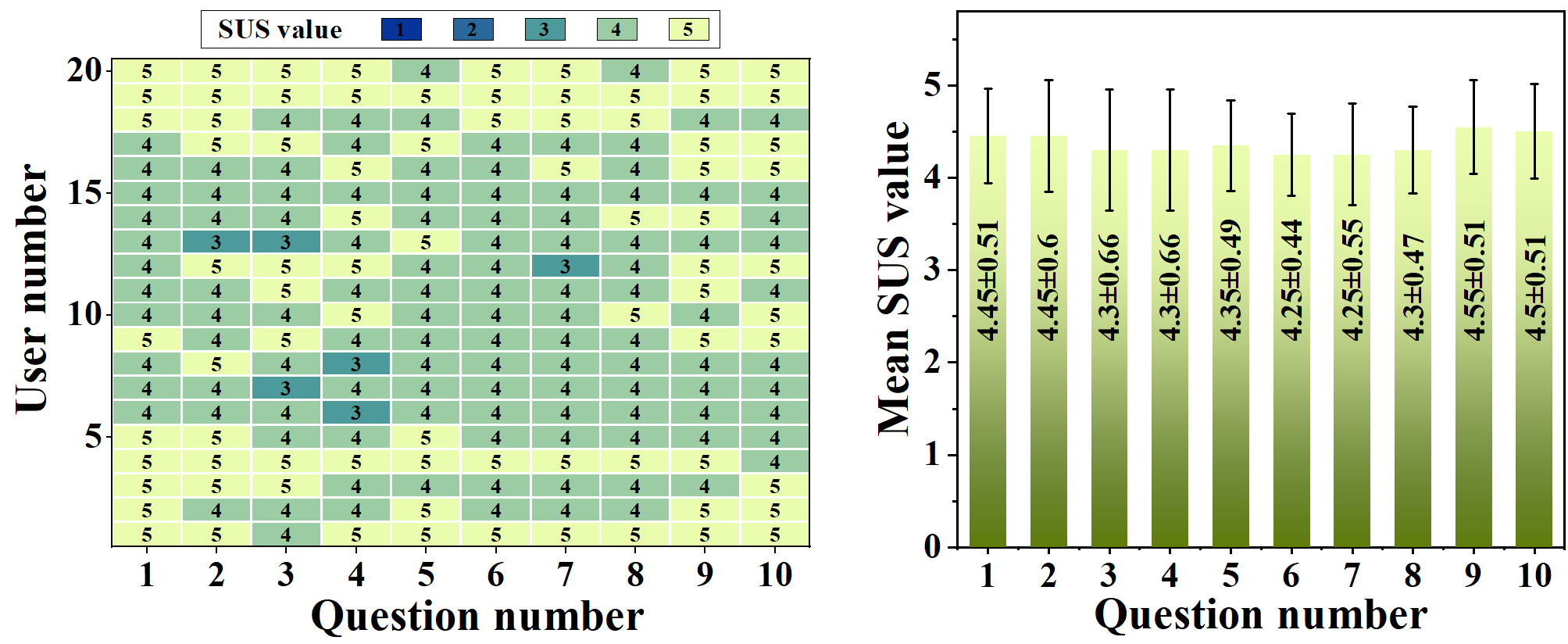}
  \caption{System Usability Scale (SUS) responses from the 20 experimental-group participants on the 10-item questionnaire: (a, left)~heatmap of each participant's individual response (rated 1--5) for each item, with cell values shown explicitly; (b, right)~bar chart of the mean SUS value per item with standard-deviation error bars and numerical labels (mean~$\pm$~standard deviation).}
  \label{fig:sus}
\end{figure}

System usability was assessed using the System Usability Scale (SUS); item-level responses (after reverse-coding for negatively worded items, as described in Section~4.4) and group-level means are summarised in Figure~\ref{fig:sus}. The 10 item-level means clustered tightly between $M = 4.25$ and $M = 4.55$ on the 1--5 response scale, with low item-level dispersion (all $SDs \le 0.66$). Items capturing voluntary engagement and self-perceived independence received the highest ratings---in particular Item~1 (\emph{``willing to play this game frequently to practice CPR''}, $M = 4.45$, $SD = 0.51$) and Item~9 (\emph{``confident in playing this game independently''}, $M = 4.55$, $SD = 0.51$). Items targeting potential complexity or learnability concerns (Items~2, 4, 6, 8, and~10), once reverse-coded, also clustered above $M = 4.25$, indicating that participants did not perceive the system as overly complex, awkward, or instruction-heavy.

Computed via the standard SUS scoring procedure, the overall score was $M_{\mathrm{SUS}} = 84.3$ (on the 0--100 SUS scale), corresponding to an \emph{Excellent} rating on Bangor et~al.'s adjective scale. Item-level responses showed within-item variability across all ten items, with each item receiving a mixture of 4s and 5s (cf.\ heatmap in Fig.~\ref{fig:sus}); the consistently high item-level ratings together with this item-by-item spread suggest that participants found TibetCPR easy to operate and felt confident engaging with the training under short, self-guided use conditions, even though the sample spanned a wide range of ages, occupations, and educational backgrounds.

Taken together with the depth-accuracy results above, the SUS findings indicate that participants' progressive performance gains were accompanied by uniformly positive usability perceptions, supporting the design intent of enabling brief, self-guided practice across a demographically diverse user base.

\subsection{Qualitative Results}

To complement the performance metrics, we conducted semi-structured interviews with the 20 experimental-group participants to understand how they experienced TibetCPR in practice, including how they interpreted system feedback, what supported continued engagement, and what aspects required adjustment during use. Our thematic analysis identified six recurring experience-level themes related to interaction and use: (1)~overall experience and learnability; (2)~engagement with narrative and progression elements; (3)~experiences with multimodal feedback; (4)~language and contextual familiarity during use; (5)~perceived skill awareness and confidence; and (6)~suggested directions for future improvement.

\subsubsection{Overall experience and learnability: ``After a few minutes I basically grasped the rules''}

Most participants described an initial sense of uncertainty when first interacting with TibetCPR, followed by a relatively rapid transition toward more confident and directed action. In their accounts, learnability centred on how to interpret and respond to system feedback during practice. As P4 reflected, ``\emph{At the very beginning I felt a little nervous---I was a bit afraid that the electric stimulation might be too uncomfortable, and there was also a kind of psychological burden about whether I could really do it well. But after a few minutes, I basically grasped the rules and the right direction for improvement.}''

Several participants described the first few minutes of interaction as a calibration period, during which they learned to associate specific feedback cues with corrective actions. P4 added, ``\emph{At the start I didn't know anything and was just pressing randomly. The system itself provided a learning process, and that mattered.}'' In the absence of instructor guidance, this early calibration phase played an important role in helping users connect feedback signals with bodily actions; the subjective trajectory participants described here closely parallels the gradual minute-by-minute improvement we observed in both rhythm and depth during T2 (cf.\ Figs.~\ref{fig:rate-trajectory}, \ref{fig:depth-trajectory}).

Beyond the calibration period, participants noted that the system reduced guesswork by making performance states perceptible during interaction. P1 explained, ``\emph{The prompts---the electric stimulation and the visual colour---were very useful. They worked like an intuitive reference, which made learning easier; the system clearly told me what was right and what wasn't.}''

Overall, participants' accounts characterised learnability as a gradual shift from initial uncertainty to a clearer understanding of how to respond to system feedback during use, with confidence emerging across the first few minutes of interaction.

\subsubsection{Engagement with narrative and progression elements}

Participants frequently discussed the snow-lotus narrative and the ``life-tree'' progression metaphor in relation to their engagement during practice. Many described these elements as emotionally resonant and helpful for sustaining attention across repeated training cycles. As P19 reflected, ``\emph{I really like the snow-lotus expression---this is something unique to Tibet, and it gives me a sense of warmth that's hard to put into words. The way the snow lotus blooms and withers also resonates with how Tibetan culture talks about the bloom and decline of life---I find that very fitting.}''

Beyond cultural resonance, participants described the game's progression structure as a meaningful indicator that their effort was accumulating over time. P16 commented, ``\emph{I like the life-tree design. I think this image is a very good fit for Tibet---in other places trees might seem ordinary, but in Tibet trees grow under extremely harsh conditions and grow very slowly. Cultivating the life tree feels a bit like watching our own homeland.}'' Several participants similarly described the visible growth of the tree as encouraging continued effort, particularly during the more repetitive parts of training. P2 added, ``\emph{Every time I save someone in the game, the life tree grows a little. It really gives me a strong sense of achievement.}''

Rather than treating these elements as external rewards, participants framed the narrative as an embodiment of accumulated practice. The narrative framing thus appeared to reinforce continued engagement during the otherwise repetitive compression task---a pattern broadly consistent with the experimental group's sustained improvement across the full 10-minute T2 phase rather than levelling off after the first few minutes.

\subsubsection{Experiences with multimodal feedback}

Participants frequently described multimodal feedback as a central aspect of how CPR practice became actionable during interaction, and their accounts mapped onto the two performance outcomes in distinct ways: visual cues were described as continuous, low-cost monitoring signals, while electrotactile cues were described as discrete, attention-grabbing error signals.

Visual cues were typically described as easy to interpret and low in cognitive demand---a characterisation consistent with their role in supporting the gradual rhythm convergence visible in Fig.~\ref{fig:rate-trajectory}(a). P10 noted, ``\emph{I could judge the rhythm by the colour, which is very simple. I really liked this kind of feedback---the scale was just right. While I was pressing, I had little spare attention for other things, but distinguishing one colour from another was within my comfort range. Honestly, it was still a little tiring.}''

Electrotactile feedback, by contrast, was commonly described as salient and difficult to ignore, especially when signalling incorrect compressions. This immediacy is consistent with its plausible role in driving the substantial depth-accuracy improvement we observed during T2 (Fig.~\ref{fig:depth-trajectory}(a)). P5 remarked, ``\emph{The electric-stimulation feedback was very responsive and very useful---if I made a mistake, the prompt came immediately, and that actually made me feel safer.}''

At the same time, participants did not experience electrotactile feedback uniformly. A small number reported some physical discomfort and raised concerns about individual differences in tolerance. P20 explained, ``\emph{If possible, could the electric stimulation be a little gentler? I still felt a bit of discomfort.}'' Beyond the physical sensation, P16 described a more psychological response: ``\emph{I was a bit scared of the electric stimulation; it brought some psychological pressure.}''

Together, these accounts suggest that while the immediacy of electrotactile feedback was valued and appears to underpin the depth-accuracy gains, intensity calibration and individual sensitivity remain important design considerations for accommodating users with different tolerance thresholds.

\subsubsection{Language and contextual familiarity during use}

Participants discussed language support as a key factor influencing the system's approachability during use. Tibetan-language prompts were frequently mentioned as lowering the barrier to independent interaction, particularly for participants whose primary language was Tibetan. P5 explained, ``\emph{I really like the Tibetan prompts. As a native Tibetan speaker, although I can understand most Mandarin, Tibetan still feels much more familiar to me.}''

P13 articulated this point through a comparative analogy: ``\emph{Language really matters. I appreciate the language support you've put in---hearing my own language helps me relax. Imagine being addressed in Mandarin versus English: if you're a native Mandarin speaker, you'd obviously feel more at ease in the former.}'' Several other participants similarly noted that receiving feedback in their primary language reduced hesitation during practice and made it easier to interpret system cues without external assistance. Beyond language, participants also described the highland-pasture visual context as contributing to a sense of familiarity, framing the system as situated within their everyday surroundings rather than a generic training tool. These accounts of contextual fit are consistent with the high SUS ratings on items related to independent and unassisted use---in particular Item~9 (\emph{``confident in playing this game independently''}, $M = 4.55$, $SD = 0.51$; cf.\ Fig.~\ref{fig:sus}).

\subsubsection{Perceived skill awareness and confidence}

Most participants reported developing a clearer sense of what correct CPR performance should feel like, especially with respect to compression depth and rhythm---a subjective shift that aligns with the substantial T1$\rightarrow$T3 gains we observed on both outcomes (cf.\ Fig.~\ref{fig:rate-mean}). P2 reflected, ``\emph{I think I learned a lot of details---roughly what frequency and force to apply. More importantly, TibetCPR made me feel that performing CPR isn't as difficult as I had imagined.}''

At the same time, participants articulated their confidence in cautious terms. Several described feeling more willing to attempt CPR than before, while also acknowledging the limits of a single short session. P15 commented, ``\emph{Through the training I feel I have already learned a lot, but there is still much I would need to practise. As things stand now, if you asked me to actually perform CPR in a real emergency, I wouldn't dare.}'' This framing of confidence as increased readiness rather than full preparedness was echoed by other participants and is consistent both with our characterisation of T3 as a measure of short-term transfer and with the design intent of TibetCPR as a low-burden, self-guided introduction to chest-compression performance rather than a substitute for hands-on instructor training.

\subsubsection{Suggested directions for future improvement}

Beyond describing their experiences, several participants offered concrete suggestions for refining the system. Three sets of suggestions emerged most frequently: feedback adjustability, breadth of training content, and pacing of initial onboarding.

On the feedback side, multiple participants echoed the desire for finer control over electrotactile-stimulation intensity to accommodate individual sensitivity differences---picking up on the same heterogeneity in tolerance that surfaced earlier in the multimodal-feedback theme above. Beyond intensity adjustment, several participants suggested enriching the auditory channel. P7 commented, ``\emph{If there were Tibetan-language encouragement voices, I would feel more at home---it shouldn't only be about correcting mistakes; it could also help me relax a little during practice.}''

On the training-content side, participants noted that the current system mainly trains the chest-compression action itself, while real-world rescue involves a wider sequence of steps. P13 commented, ``\emph{Right now I feel I'm mainly practising the pressing motion, but a real rescue has many other steps. I hope future versions of the game can include more of the full process.}''

On the onboarding side, some participants suggested introducing information at a slower pace in the early stages to ease the comprehension load. P4 noted, ``\emph{If at the very beginning it could teach a bit more slowly, step by step, I think it would be even easier to adapt.}''

Together, these suggestions point toward three concrete directions for future iterations of the system---adjustable feedback intensity, expanded coverage of the full CPR procedure, and a more gradual onboarding curve---that participants saw as natural next steps from their lived experience.

%% file: sections/Discussion.tex
\subsection{From Immediate Correction to Gradual Stabilization}

Across both rhythm and depth measures, performance improvements associated with TibetCPR did not emerge immediately upon exposure to feedback, but accumulated progressively over the course of interaction. A linear mixed-effects model revealed a robust Group$\times$Minute pattern for both outcomes: the experimental group's rate deviation decreased approximately 2.5 units per minute faster than the control group's ($\beta = -2.52$ per minute, $p < .001$), and depth accuracy rose roughly three percentage points per minute faster ($\beta = +0.030$ per minute, $p < .001$). At minute~1 of training, the two groups did not differ significantly on either outcome; by minute~10, the gaps were large and statistically clear. This temporal pattern stands in contrast to feedback paradigms organised around instantaneous error correction, where a single trial's correction is expected to produce immediate compliance \cite{song2016smartwatches}.

Three features of the data converge on reading this trajectory as gradual recalibration. First, improvements in the experimental group accumulated incrementally across all ten minutes of T2 rather than being concentrated at any single point. Second, participants in the qualitative interviews described the first few minutes as a calibration period---a subjective characterisation that closely parallels the model-estimated trajectory. Third, the control group, who completed identical practice on the same mannequin without feedback, showed no comparable trend on rhythm and only a small within-session practice effect on depth (paired $d_z = 0.53$) that was approximately seven times smaller than the experimental gain and did not carry through to the unscaffolded post-test. Together, these patterns are consistent with motor-learning accounts in which feedback functions as a reference signal supporting iterative self-adjustment rather than as a prescriptive corrector \cite{abella2005quality}.

The fact that experimental-group performance also transferred to an unscaffolded one-minute post-test (T3), conducted with feedback removed, suggests that the observed improvements were not contingent on the moment-to-moment presence of system support. Because T3 immediately followed T2 with no washout period, this transfer should be understood specifically as \emph{short-term motor performance transfer to an unscaffolded condition} \cite{kramer2006quality, meng2026engagement}; assessing performance maintenance over hours, days, or weeks would require delayed post-tests, which our single-session protocol could not accommodate. We return to this temporal scope under limitations.

Within the experimental group, the magnitude of improvement was larger for participants who began with less consistent baselines, suggesting that the system's effects were not concentrated among already-strong performers. This observation is exploratory---our study was not designed as a moderator analysis---but it points to a direction worth examining in larger samples, and is potentially valuable in self-guided deployments serving demographically heterogeneous users with varying physical preparation.

\subsection{Multimodal Feedback as a Calibration Ecology}

The gradual trajectory described above invites closer attention to \emph{which} feedback channels did the work, and \emph{how} they did it. Our results suggest that the visual and electrotactile channels in TibetCPR contributed in complementary rather than redundant ways, with each channel's temporal granularity aligned to the temporal structure of the outcome it most plausibly supported.

Visual feedback, conveyed through colour-coded state indicators, was experienced by participants as low in cognitive demand and easily monitored at a glance during physically demanding action. Quantitatively, this characterisation is consistent with the trajectory of rhythm stabilisation: rate deviation decreased smoothly and continuously over minutes (Fig.~\ref{fig:rate-trajectory}(a)), the kind of pattern that would be expected when a continuously available cue supports continuous correction over a continuous control variable. Such low-cost, glance-able encoding of state has been associated in prior MobileHCI work with sustained motor performance under attention competition \cite{swain2024assessing, lu2020wearable}.

Electrotactile feedback, by contrast, was described as salient, episodic, and tightly coupled to perceived errors. Although the depth-accuracy trajectory itself rose smoothly across the session (Fig.~\ref{fig:depth-trajectory}(a)), participants' qualitative accounts characterise this rise as driven by attending to discrete tactile alerts on out-of-range compressions rather than by continuous monitoring of a state variable \cite{lee2024soft}. Taken together with the variables' own structure---compression rhythm is naturally a continuous control variable, whereas compression depth is naturally a categorical in-range/out-of-range judgement---the two modalities can be read as mapping onto distinct \emph{temporal granularities} of correction (continuous monitoring versus discrete error signalling) that align with the temporal structure of the behaviour each was paired with. The system's efficacy may rest in part on this alignment between feedback temporal granularity and the temporal structure of the target behaviour.

This framing also helps interpret the heterogeneity participants reported in their experience of electrotactile feedback. While most participants found the immediacy of electric stimulation valuable, a subset reported physical discomfort or anticipatory unease, and several called for adjustable intensity. These reports highlight that high-salience tactile cues are perceptually and emotionally consequential, particularly for users with limited prior exposure to such systems \cite{lin2025use}. Sensitivity-aware intensity calibration is a near-term design step, and we return to this point under future directions.

Beyond the specifics of CPR, this analysis suggests a more general design heuristic for self-guided embodied training: rather than maximising modality count, designers may benefit from matching modality temporal granularity to the temporal structure of the target behaviour---using continuous, low-cost channels for continuous control variables, and discrete, salient channels for categorical judgements. We develop this point further in Section 6.5.

\subsection{Autonomous Interpretability as a Deployment Prerequisite}

A consistent feature of the present study, by design, is that participants received only a brief orientation to the system before practice began. No instructor mediated their use, no demonstration accompanied the feedback signals, and the participants themselves spanned a wide range of ages, occupations, and educational backgrounds---including participants with limited literacy. That participants nonetheless reached substantial post-training performance, and reported high overall confidence in independent system use (overall SUS = 84.3 on a 0--100 scale, on Bangor et~al.'s adjective scale corresponding to an \emph{Excellent} rating), points to a design quality that we characterise as \emph{autonomous interpretability}: the property of a feedback signal that allows it to be understood and acted upon through the user's own bodily action, without external mediation.

This formulation differs from the usability-as-simplicity framing common in self-guided systems. Participants' qualitative accounts of learnability did not focus on the number of interface elements or on minimising menus; they focused on the few minutes during which they came to associate specific feedback cues with specific bodily corrections (cf.\ Section 5.3.1). What made the system tractable was not the absence of complexity but the legibility of its evaluative signals---in our case, the binding of colour to rhythm and of stimulation to depth, and the immediacy with which both were tied to the user's physical action. In this respect our findings echo prior work on self-directed motor learning and ambient interaction in which feedback that is interpreted through the user's own activity, rather than read off a separate display, supports independent practice \cite{jeon2021effectiveness, sevil2021effect, song2016smartwatches}.

The relevance of this distinction is not academic. In settings such as the one we studied---where instructor support is fragmented or absent, where users vary widely in literacy, and where training opportunities are short and opportunistic---autonomous interpretability is not a desirable feature but a deployment prerequisite. A system whose feedback requires explanatory mediation will not, in such contexts, be used on its own terms. The SUS data are consistent with this reading: item-level responses showed variability across all ten items (cf.\ Section 5.2), suggesting that participants engaged with the distinct usability features each item targets rather than registering an undifferentiated overall impression.

We do not claim that minimal instruction is desirable in all training scenarios; for many forms of clinical CPR practice, instructor-led teaching remains the standard and should remain so. We do claim that, in deployment contexts where instruction cannot be assumed, the question of whether feedback is autonomously interpretable is prior to questions about feedback richness or interface aesthetics. We return to this in Section 6.5.

\subsection{Localisation as Contextual Fit, Not a Learning Mechanism}

A separate but related design dimension of TibetCPR was its localisation: Tibetan-language prompts, visual elements drawn from highland environments, and a narrative built around the snow lotus and a growing life tree. Participants spoke about these elements warmly and repeatedly. Our quantitative measures, however, do not show---and were not designed to show---that localisation directly accelerated learning or improved compression accuracy. We did not run a non-localised arm. The contribution we draw from this section is therefore more careful than a causal claim.

Participants' accounts suggest that localisation primarily shaped \emph{whether and how} they engaged with the system, not the rate at which their motor performance changed. Tibetan-language prompts were repeatedly described as reducing hesitation during self-guided practice (cf.\ Section 5.3.4)---not because Mandarin was unintelligible to most participants, but because the affective experience of receiving instruction in one's primary language lowers the threshold to acting on it. Cultural elements such as the snow lotus and the life tree were similarly described as contributing to a sense that the system belonged in participants' surroundings. These accounts align with cultural and ICT-for-development HCI work emphasising that ostensibly universal designs carry cultural defaults whose absence-of-fit, rather than presence-of-bug, can obstruct adoption \cite{chen2017public, teng2020awareness, zhao2025immersive, zeng2025parental}.

Our statistical claim is therefore narrow: in our data, the gains in rhythm and depth are attributable to the feedback dynamics characterised in Sections 6.1 and 6.2, not to the localised wrapping in which they were delivered. Our design claim is broader: localisation appears to have shaped \emph{whether participants chose to engage with the system in a self-guided manner at all}, which is a precondition for the feedback dynamics to do any work. We read this as a separation of concerns---learning mechanism versus contextual fit---rather than as a hierarchy.

This separation has practical implications for mobile health training systems intended for deployment beyond the contexts in which they were designed \cite{cao2026causalinfluencemaximizationsteadystate,liang2026miraembeddingsv1domainadaptedsemanticreranking}. When localisation is treated as a learning intervention, evaluations that fail to find direct skill-acquisition effects can be misread as evidence that cultural adaptation does not matter. When localisation is treated as a contextual condition for engagement, the same evaluations can be read as evidence that engagement scaffolding is doing exactly what it should: setting the stage on which interaction takes place. The lay community sample we recruited---spanning a wide range of ages, occupations, and educational backgrounds---makes the willingness-to-engage frame more, not less, important. In samples this heterogeneous, willingness is what brings users to the interaction in the first place.

\subsection{Synthesis: Three Design Principles for Self-Guided Embodied Skill Training}

We close the substantive Discussion by stepping back from the specifics of CPR and Tibet to ask what TibetCPR's design choices, viewed through the present results, might contribute as transferable design knowledge for self-guided embodied skill training systems more broadly. We see three principles emerging.

\textbf{P1. Feedback as a calibration reference, not an immediate corrector.} The trajectory data and the qualitative accounts of a brief calibration period both suggest that effective feedback in self-guided training need not---and perhaps should not---enforce correctness in the moment. Treating feedback as a continuously available reference against which the user's bodily action is compared, rather than as a directive to be obeyed, supports the iterative recalibration that produced the bulk of the gains we observed. This design move is consistent with the long-standing motor-learning observation that overly prescriptive feedback can foster dependence on the feedback signal itself \cite{abella2005quality}.

\textbf{P2. Match modality temporal granularity to the temporal structure of the target behaviour.} Compression rhythm is a continuous variable and benefited from continuous, low-cost visual encoding. Compression depth involves a categorical in-range/out-of-range judgement and benefited from discrete, salient electrotactile signalling. The design heuristic that follows is concrete: before adding modalities, characterise the temporal structure of each control variable in the target task---continuous, episodic, or compositional---and select modality temporal granularity accordingly. We see this as a more disciplined alternative to the implicit assumption that more modalities are better.

\textbf{P3. Autonomous interpretability is the precondition for deployment, not the after-effect of usability.} Where instructor mediation cannot be assumed, the question of whether feedback can be understood through one's own action is prior to questions about interface complexity, aesthetic minimalism, or task scaffolding. We have characterised this as a property of the feedback design itself, not merely an attribute of the interaction surface, and we suggest it deserves first-order treatment in design processes for self-guided systems.

These three principles are framed at a level of generality that, we hope, abstracts away from CPR. They are also intentionally modest. None alone is sufficient for a successful self-guided embodied training system; together, they describe a small set of design moves that the TibetCPR data suggest are productively combinable. We expect each principle to interact with task-specific and culturally-specific considerations, and we offer them as starting points for further empirical examination rather than as design rules.

\subsection{Limitations and Future Directions}

This study has several limitations that bound our claims. First, the evaluation was a single-session design with the post-test (T3) administered immediately after the intervention and no washout period. The transfer we observed is therefore short-term and unscaffolded, not long-term retention; we do not know whether the patterns we observed would survive a delay of hours, days, or weeks, nor how brief refresher sessions might support consolidation. Longitudinal evaluation with delayed post-tests, possibly paired with spaced refresher interactions, is a natural next step.

Second, our control condition was unguided practice on the same mannequin without feedback, which is an appropriate baseline for the interaction-level question we asked but does not position TibetCPR against existing CPR feedback systems (e.g., commercial QCPR-class devices) or against instructor-led training. We therefore do not make claims about relative effectiveness; a three-arm comparison would be required for such claims.

Third, although our sample of lay community members spans a wide range of ages, occupations, and educational backgrounds---broadening the external relevance of our claims about autonomous interpretability and contextual fit---all participants resided in the Tibetan Autonomous Region. Whether the design principles articulated in Section 6.5 transfer to other high-altitude or low-resource settings remains to be tested. The system also covers only chest-compression rhythm and depth and does not currently support scene assessment, emergency communication, or AED use; participants identified this as a gap (cf.\ Section 5.3.6), and whether the interaction principles scale to multi-step CPR workflows is an open empirical question.

Fourth, because our intervention always combined visual and electrotactile feedback, we cannot decompose their relative contributions experimentally. Our claim about complementary temporal granularities (Section 6.2) rests on triangulating quantitative trajectory shapes with qualitative accounts rather than on direct manipulation, and a modality-ablation study is a clear next step.

Fifth, the design choices reported in Section 3.1 rest on prior motor-learning and CPR-feedback literature together with contextual reasoning about the deployment setting, rather than on a formative user study or pre-deployment co-design with the local target population. While we believe the literature-anchored rationale is defensible for an initial deployment, future iterations would benefit from participatory design with local stakeholders to validate context-specific assumptions before scaling \cite{he2023exploring, chen2023design, zeng2025parental}.

Finally, the post-test effect sizes we observed were large, and effect sizes of this magnitude in compact randomised designs warrant caution. Because the electrotactile feedback is perceptually salient, participants in the experimental group were necessarily unblinded to their condition; although the pattern of progressive gains across T2 and the transfer to an unscaffolded T3 is difficult to reconcile with a pure demand-characteristics explanation, we do not rule out that some component of the effect reflects participants' awareness of the study's purpose. Replication in naturalistic deployment would address this directly.

%% file: sections/Conclusion.tex
This paper presented \textit{TibetCPR}, a low-cost, self-guided CPR training system designed for autonomous use in high-altitude, resource-constrained settings. Through a randomised study with 40 lay community members in the Tibet Autonomous Region, we examined whether brief, game-based interaction with depth-driven electrotactile feedback and rhythm-driven visual cues could stabilise chest compression rhythm and depth without instructor mediation.

Performance gains accumulated progressively across the 10-minute intervention rather than at first exposure, and transferred to an unscaffolded one-minute post-test conducted with feedback removed. Qualitative accounts and high overall usability ratings (SUS = 84.3) further suggest that participants found the system interpretable and operable on their own, even across a sample spanning ages 19--56 and a wide range of occupations and educational backgrounds. We synthesise these findings into three design principles for self-guided embodied skill training: feedback as a calibration reference rather than an immediate corrector, modality--outcome alignment by temporal granularity, and autonomous interpretability as a deployment prerequisite rather than an after-effect of usability. By positioning Tibet not as a deployment backdrop but as a context that makes these constraints explicit, this work contributes interaction-level design knowledge for mobile, self-guided health training systems beyond institutional settings.